\documentclass[twocolumn,twocolappendix]{aastex63}
\graphicspath{{./Figs/}{./png/}}
\usepackage{amsmath}
\usepackage{multirow}

\newcommand{\K}{\,{\rm K}}
\newcommand{\G}{\,{\rm G}}
\newcommand{\nG}{\,{\rm nG}}
\newcommand{\g}{\,{\rm g}}
\newcommand{\cm}{\,{\rm cm}}
\newcommand{\kpc}{\,{\rm kpc}}
\newcommand{\Mpc}{\,{\rm Mpc}}
\newcommand{\MeV}{\,{\rm MeV}}
\newcommand{\GeV}{\,{\rm GeV}}
\newcommand{\Sec}[1]{Section~\ref{#1}}

\usepackage{scalerel}
\newcommand\HI{H\protect\scaleto{$I$}{1.2ex}}

\shorttitle{Magnetic fields during structure formation}
\shortauthors{Mtchedlidze et al.}
\graphicspath{{./}{figures/}}

\begin{document}

\title{Evolution of Primordial Magnetic Fields during Large-scale Structure Formation}

\author{Salome Mtchedlidze}
\affiliation{School of Natural Sciences and Medicine, Ilia State University, 3-5 Cholokashvili St., 0194 Tbilisi, Georgia}
\affiliation{Institut f\"ur Astrophysik, Georg-August-Universit\"at G\"ottingen, Friedrich-Hund-Platz 1, D-37077 G\"ottingen, Germany}
\affiliation{Abastumani Astrophysical Observatory, Tbilisi, GE-0179, Georgia}
\author{Paola Dom\'inguez-Fern\'andez}
\affiliation{Hamburger Sternwarte, Universit\"at Hamburg, Gojenbergsweg 112, 21029 Hamburg, Germany}
\affiliation{Department of Physics, School of Natural Sciences UNIST, Ulsan 44919, Republic of Korea}

\author{Xiaolong Du}
\affiliation{Carnegie Observatories, 813 Santa Barbara Street, Pasadena, CA 91101, USA}

\author{Axel Brandenburg}
\affiliation{Nordita, KTH Royal Institute of Technology and Stockholm University, Hannes Alfv\'ens v\"ag 12, SE-10691 Stockholm, Sweden}
\affiliation{The Oskar Klein Centre, Department of Astronomy, Stockholm University, AlbaNova, SE-10691 Stockholm, Sweden}
\affiliation{McWilliams Center for Cosmology and Department of Physics, Carnegie Mellon University, 5000 Forbes Ave, Pittsburgh, PA 15213, USA}
\affiliation{School of Natural Sciences and Medicine, Ilia State University, 3-5 Cholokashvili St., 0194 Tbilisi, Georgia}

\author{Tina Kahniashvili}
\affiliation{McWilliams Center for Cosmology and Department of Physics, Carnegie Mellon University, 5000 Forbes Ave, Pittsburgh, PA 15213, USA}
\affiliation{School of Natural Sciences and Medicine, Ilia State University, 3-5 Cholokashvili St., 0194 Tbilisi, Georgia}
\affiliation{Abastumani Astrophysical Observatory, Tbilisi, GE-0179, Georgia}

\affiliation{Department of Physics, Laurentian University, Ramsey Lake Road, Sudbury, ON P3E 2C, Canada}

\author{Shane O'Sullivan}
\affiliation{School of Physical Sciences and Centre for Astrophysics \& Relativity, Dublin City University, Glasnevin, D09 W6Y4, Ireland}

\author{Wolfram Schmidt}
\affiliation{Hamburger Sternwarte, Universit\"at Hamburg, Gojenbergsweg 112, 21029 Hamburg, Germany}

\author{Marcus Br\"uggen}
\affiliation{Hamburger Sternwarte, Universit\"at Hamburg, Gojenbergsweg 112, 21029 Hamburg, Germany}

\begin{abstract}

Primordial magnetic fields (PMFs) could explain the large-scale 
magnetic fields present in the universe.
Inflation and phase transitions in the early universe could
give rise to such fields with unique characteristics.
We investigate the magnetohydrodynamic evolution of these
magnetogenesis scenarios with cosmological simulations.
We evolve inflation-generated magnetic fields either as
(i) uniform (homogeneous) or as
(ii) scale-invariant stochastic fields, and phase-transition-generated ones either as
(iii) helical or as (iv) nonhelical fields from
the radiation-dominated epoch.
We find that the final distribution of magnetic fields in the
simulated cosmic web shows a dependence on the initial strength
and the topology of the seed field.
Thus, the observed field configuration retains information on the initial conditions 
at the moment of the field generation. 
If detected, PMF observations would open a new window for indirect probes of the early universe.
The differences between the competing models are revealed on the
scale of galaxy clusters, bridges, as well as filaments and voids.
The distinctive spectral evolution of different seed fields produces
imprints on the correlation length today.
We discuss how the differences between rotation measures from highly
ionized regions can potentially be probed with forthcoming surveys. 
\end{abstract}

\keywords{Large scale structure of universe, primordial magnetic fields, numerical simulations}

\section{Introduction} \label{sec:intro}

Magnetic fields pervade the universe on different observable scales
from planets and stars \citep[e.g.,][]{Stevenson2010,Schubert2011} to galaxies
and galaxy clusters  \citep[e.g.,][]{Becketal2001,BeckWielebinski2013}.
Different observational techniques reveal the large-scale morphology of
these fields on galaxy and galaxy-cluster scales with microgauss strengths and correlation lengths reaching a few tens of kiloparsec in galaxy clusters \citep[see, e.g.,][]{Murgiaetal2004,Vogt2005}.

The existing theories for explaining such large-scale
correlated magnetic fields can be broadly divided into two classes:
(1) the primordial scenario, where a seed 
magnetic field is generated in the early universe, during epochs preceding the structure formation \citep{Kandusetal2011}; and (2)
the astrophysical scenario, where the observed magnetic fields
have their origin in an initial weak seed field produced in
astrophysical sources (e.g., stars or active galactic nuclei within (proto-)galaxies) due to local mechanisms (e.g., Biermann battery; see \citealt{Biermann1950}), and then amplified and transferred to larger scales
 \citep{Kulsrudetal2007,Ryuetal2008}. 
For instance, the Biermann battery mechanism can generate seed magnetic fields from an initially zero magnetic field 
in cosmological shocks driven by the gravitational structure formation
\citep{1998A&A...335...19R} or in ionization fronts during the Epoch of
Reionization \citep{Subramanianetal1994,2000ApJ...542..535G}.\footnote{The
reader may also refer to the following recent works:
\citealt{NaozNarayan2013,Garaldietal2021}, and \citealt{Attiaetal2021}.}
Yet, it remains unclear how efficiently magnetic field seeds can
spread in the astrophysical scenario 
(see, e.g., \citealt{Bertoneetal2006} and \citealt{Marinnacietal2015} for studies on galactic winds).

Studies of Faraday rotation measurements and MgII absorption lines
(tracing halos of galaxies) have shown that microgauss magnetic fields
were already present in Milky Way-type galaxies a few billion years after the
Big Bang \citep[e.g.,][]{Kronberg2008,Bernetietal2008,Bernetietal2010}.
Moreover, observations of blazar spectra by the Fermi Gamma Ray
Observatory indicate that cosmic voids can host weak $10^{-16}$--$10^{-15}\,$G 
magnetic fields (coherent on megaparsec scales) which favors a primordial scenario;
\footnote{Blazar spectra measurements can also be used to constrain the volume-filling factor of extragalactic magnetic fields seeded by starburst galaxies or active galactic nuclei; see \citep{Dolagetal2011} } see, e.g., \citealt{NeronovVovk2010,Tavecchioetal2010}
for pioneering studies and \cite{Vachaspati2020} for a review and references therein; also see \cite{Arlen:2012iy}
for discussions of possible uncertainties in the measurements of blazar
spectra, and \cite{Broderick:2018nqf} and \cite{AlvesBatista:2019ipr}
on the possible impacts of plasma instabilities.

Seed magnetic fields generated in the early universe (e.g.,
primordial magnetic fields, PMFs) through its stress–energy
tensor induce scalar (density), vector (vorticity), and
tensor (gravitational waves) perturbations 
(see \citealt{Subramanian2016} for a review) 
and through these
perturbations may leave potentially observable traces.
In particular, the PMF effects include additional large- and small-scale angular anisotropies in the cosmic microwave background (CMB) temperature and polarization
\citep[see][and references therein]{Plancketal2016}, and changes
in the matter power spectrum 
(see \citealt{Wasserman1978,Kimetal1996} and \citealt{GopalSethi2003} for seminal work;
and \citealt{FedeliMoscardini2012,Kahniashvilietal_2013_2, Sanatietal2020} and  \citealt{Katzetal2021} 
for recent studies). Furthermore, the dissipation of PMFs throughout the 
intergalactic medium (IGM) in the pre-recombination era leads to spectral 
distortions in the CMB \citep{Jedamziketal2000}, while their dissipation in the
post-recombination era changes the thermal state of the IGM  and alters the
reionization history of the universe \citep{Jedamzik1998,SethiSub2005,SethiSub2009}.
The observational signatures of PMFs also include the effects on the CMB
polarization plane rotation, the Faraday rotation \citep[CMB
birefringence; see e.g.,][]{POLARBEAR2015,MinamiKomatsu2020} 
and the effects on light-element
abundances during the big bang nucleosynthesis (BBN); see, e.g.,
\cite{YamazakiKusakabe2012} and \cite{Louetal2019}. 
At high redshifts, PMFs are proposed to lower the angular momentum barrier of a collapsing gas leading to direct collapse black hole (DCBH) formation \citep[][]{Pandeyetal_2019}.
Interestingly, recently PMFs have also been considered as a potential possibility 
to relax the Hubble tension \citep[][]{Jedamzik:2020krr}.

PMFs could come from different generation scenarios,
including inflation and phase transitions\footnote{Generally, the topological defects arising from
various phase transitions in the early universe could also source the
generation of PMFs; see, e.g., \cite{Horiguchietal2015} and a review
on topological defects by \cite{Durreretal2002}.
In addition, cosmic strings are another possible source of the PMF generation
\citep[see][for a review]{Vachaspati2020}.
} \citep[see][for
reviews]{GrassoRubinstein2001,DurrerNeronov2013,Subramanian2016}.
Inflationary magnetogenesis assumes that vacuum fluctuations of an
electromagnetic (or hypermagnetic) field
give rise to a weak seed field which then grows (see \citealt{TurnerWidrow1988} and \citealt{Ratra1992} for pioneering work,
and \citealt{Sharma+17,Sharma+18, Bamba:2020qdj} and \citealt{Maity:2021qps}
for recent studies 
and references therein).
This mechanism usually involves the breaking of the conformal invariance
of the space-time \citep[see, e.g.,][]{Dolgov1993}.
The correlation lengths of inflation-generated magnetic fields are not
limited by causality requirements due to rapid exponential stretching
during inflation. On the other hand, a phase-transition mechanism
implies that the coherence scales of the PMFs are bound
due to causality by the Hubble horizon length scale at
the moment of field generation, i.e., electroweak (EW, $T\sim100\GeV$)
or quantum-chromodynamical (QCD, $T\sim150\MeV$) phase transitions;
see \cite{Hogan1983,Quashnocketal1989,Vachaspati1991} and \cite{Tajimaetal1992} 
for pioneering work, and \cite{Subramanian2016} and \cite{Vachaspati2020} for reviews. 
The magnetic fields from phase transitions are assumed to have a stochastic
distribution with either a power-law or a sharply peaked magnetic power 
spectrum (where the peak or the characteristic length scale is set by 
the phase transition bubble size; see, e.g., \citealt{Kahniashvilietal2010}), 
while the inflationary mechanism can produce a stochastic magnetic field 
with a scale-invariant spectrum or a spatially uniform, homogeneous 
magnetic field (i.e., the \citealt{Mukohyama2016} model; see \citealt{Mandaletal2020}, and
references therein). 
Both generation scenarios, inflationary and phase-transitional, could generate helical PMFs 
(see the seminal work of \citealt{Cornwall1997}, and \citealt{Vachaspati2020}
for a review and references therein).
The relevance of a primordial, helical magnetic field is that, if ever 
detected, it will be a direct indication of parity (mirror symmetry) 
violation in the early universe and can in turn explain the 
matter–antimatter asymmetry (see \citealt{Giovannini:1997gp} and \citealt{Vachaspati2001} 
for pioneering work, and the recent work of \citealt{Fujita2016} and \citealt{Kushwaha:2021csq}).

The current observational upper limits on the strength of PMFs 
are derived from CMB observations and depend on the specific 
effects that are analyzed.
The effects of PMFs on the CMB angular power spectrum suggest a few nanogauss
for the upper bound of the amplitude of stochastic PMFs \citep[$\sim 4.4\,$nG assuming zero
helicity and $\sim 5.6\,$nG for a maximally helical field at a scale 
of $1\Mpc$ and with a power-law power spectrum;][]{Plancketal2016}. 
The strongest upper limits on the PMF arise from the inhomogeneous 
recombination induced by the small baryonic density perturbations 
\citep[][]{Jedamzik:2018itu}.
On the other hand, the nonobservation of
gigaelectronvolt photons from tera-electronvolt blazars puts a lower limit of order 
$10^{-15}$--$10^{-16}\,$G on the strength of the
primordial magnetic seed characterized by a monochromatic spectrum,
i.e., characterized by a fixed correlation length of $1\Mpc$
\citep{Tayloretal2011}.

PMFs are expected to evolve in a distinguishable fashion across
different cosmological epochs.
Their evolution has been widely studied in numerical
simulations either with pure magnetohydrodynamical
(MHD) codes (only for radiation-dominated epochs); see,
e.g., \cite{PFL76}, \cite{Christenssonetal2001}, \cite{BanerjeeJedamzik2004}, \cite{Kahniashvilietal2016},
and \cite{Brandenburgetal2018}; or, more recently, with cosmological
MHD simulations (post-recombination epoch); see,
e.g., \cite{Miniatietal2015}, \cite{Beresnyaketal2016},
\cite{Vazzaetal2017},\cite{Vazzaetal2018}, and also \cite{Donnertetal2018}
for a recent review.
The evolution of PMFs in the radiation-dominated epoch, generated
via the inflation and phase-transition
mechanisms, proceeds in a highly turbulent regime
due to the strong coupling between magnetic fields and plasma motions
\citep{BrandenburgNordlund2011}.
It has been shown that freely decaying MHD turbulence with helicity
could source energy cascades from small to larger scales and a
slowdown of the decay \citep{Christenssonetal2001, BanerjeeJedamzik2004,
Kahniashvilietal2010, BranKahn2017} leading to $10^{-15}$--$10^{-9}\,$G
magnetic fields by the end of recombination \citep{Kahniashvili2020}.
On the other hand, it is important to connect this work with
cosmological simulations, which can reproduce the observed magnetization of the universe
on larger scales ($\sim$megaparsec scales) 
and later stages (during the structure formation),
assuming either a primordial
\citep{Dolagetal1999b,Brueggenetal2005_1,Vazzaetal2014,Marinnacietal2015}
or an astrophysical scenario with various mechanisms of magnetic
seed transport \citep[e.g., star formation and outflows from active galactic nuclei,][]
{Bertoneetal2006,Donnertetal2009,Xuetal2011,Vazzaetal2017}. 
Regardless of their origin, the physical mechanisms that further amplify magnetic fields 
during cosmological structure formation are adiabatic contraction and turbulent dynamo. The latter is the preferred 
mechanism for explaining the efficient amplification of magnetic fields on 
galaxy \citep[see, e.g.,][]{Pakmor2017,RiederTeyssier2017,Steinwandeletal2020} 
and galaxy-cluster scales \citep[see, e.g.,][]{Xuetal2009,Vazzaetal2018,Dominguezetal2019,Steinwandeletal2021}. 
Nevertheless, distinguishing between different 
magnetogenesis scenarios in the high-density regions of the cosmic web could be complicated because
the memory of the seed field is believed to be erased by the turbulent dynamo
\citep[][]{2014ApJ...797..133C}\footnote{This is also the case in galaxies \citep[see, e.g.,][for cosmological simulations of galaxies]{2014ApJ...783L..20P,Marinnacietal2015,2018MNRAS.473.4077P,Garciaetal2021} }. 
This picture can change in the more rarefied regions of the cosmic web, such as cluster outskirts, filaments and
voids, where competing models can be observationally tested \citep[see, e.g.,][]{Donnertetal2009,Vazzaetal2017}.

In the present paper, the cosmological MHD simulations are performed with the
\texttt{Enzo} code \citep{Bryanetal2014}, where the initial magnetic field seeding\footnote{
Note that in general ``seeding" refers to very weak magnetic fields at the moment of generation.
However, hereinafter it will also refer to our initial conditions,
where the magnetic field may not be very weak.}
is consistent with different primordial magnetogenesis scenarios.
For the first time, we compare the large-scale evolution of inflationary and
phase-transitional magnetogenesis scenarios. We focused on studying two
subcases for each scenario.
In the inflationary case, we study (i) a uniform, homogeneous and (ii) a stochastic,
scale-invariant initial magnetic field.
In the phase-transitional case, we study (iii) a helical and
(iv) a nonhelical initial magnetic field.

This paper is organized as follows. In Section~\ref{sec:simulations} we describe the numerical scheme and initial conditions adopted in our work. In Section~\ref{sec:physMod} we discuss the physical model, define the spectral characteristics of each magnetogenesis scenario, and discuss the physical motivations.  
In Section~\ref{sec:Results} we present our results, with caveats discussed in Section~\ref{sec:NumAsp}, and we summarize our work in Section~\ref{sec:Summ}.

\begin{deluxetable*}{c| c c c c c c c}
\label{tab:Tab1}
\tablecaption{
Initial conditions for the magnetic field; 
the characteristic wavenumber and scale of the magnetic spectra is denoted by $k_{\rm peak}$ and  $\lambda_{\rm peak}$ accordingly, and $\langle B_{0}^2 \rangle$ and $\langle B_{0} \rangle$
are the mean of the initial magnetic field energy and the initial magnetic field strength respectively\tablenotemark{a}.
}
\tablehead{
\colhead{Scenario} & \colhead{Model}& \colhead{Normalization} &  \colhead{Simulation ID}& \colhead{$\langle B_{0}^2 \rangle$}  & \colhead{$\langle B_{0} \rangle$} &  \colhead{$k_{\rm peak}$}&\colhead{$\lambda_{\rm peak}$} \\
\colhead{} &\colhead{} &\colhead{(nG)}&\colhead{~}  & \colhead{ ((nG)$^2$)}  &\colhead{(nG)} &\colhead{($h\Mpc^{-1}$)} & \colhead{($h^{-1}\Mpc$)} 
}
\startdata
\multirow{6}{*}{Inflationary} & \multirow{3}{*}{(i) Uniform} & 0.1 &  u01  & 0.0072 & 0.085 &---&---\\
            &                        & 0.5 &   u05 & 0.180  & 0.390 &---&---\\
            &                        & 1   &    u1  & 0.719  & 0.804 &---&---\\
            \cline{2-8}
            & \multirow{3}{*}{(ii) Scale-invariant} & 0.1 & s01 & 0.0072 & 0.079 & 0.02 &45\\
            &                        & 0.5 & s05 & 0.180  & 0.395& 0.02 & 45\\
            &                        & 1& s1     & 0.719  & 0.790 & 0.02 & 45\\
\hline
\multirow{6}{*}{ Phase-transitional} & \multirow{3}{*}{(iii)Helical} & 0.1 &  h01  & 0.0072 & 0.080 & 0.4 & 2.6  \\
            &                            & 0.5 &   h05 & 0.180  & 0.402 & 0.4 & 2.6  \\
            &                            & 1   &    h1  & 0.719  & 0.781 & 0.4 & 2.6 \\
            \cline{2-8}
            & \multirow{3}{*}{(iv) Nonhelical} & 0.1  & nh01  & 0.0072 & 0.085 & 0.8 & 1.26\\
            &                              & 0.5  & nh05  & 0.180  & 0.424 & 0.8 & 1.26\\
            &                              & 1    & nh1    & 0.719  & 0.848 & 0.8 & 1.26\\
\enddata
\tablenotetext{a}{We use comoving quantities everywhere unless stated otherwise.}
\end{deluxetable*}
%
\section{Simulations} \label{sec:simulations}

We use the cosmological Eulerian MHD code \texttt{Enzo}
\citep{Bryanetal2014} for simulating a 
comoving volume of $(67.7 h^{-1}\Mpc)^3$ with a resolution of 132 $h^{-1}\kpc$.
We employed a static uniform grid of $512^3$ cells with $512^3$ dark matter (DM)
particles, each of mass $m_{\text{DM}} = 2.53\times 10^{8} M_{\odot}$.
We use the Dedner formulation of MHD equations that handles the
divergence-free condition with a cleaning algorithm for magnetic monopoles
\citep[][]{Dedneretal2002}.
The spatial reconstruction is done using the piecewise linear method \citep[PLM;][]{1979JCoPh..32..101V,COLELLA1985264} and fluxes at cell interfaces are calculated using the Harten–Lax–van Leer (HLL) Riemann solver \citep[][]{Toro1997}. The time integration is performed using the total variation diminishing (TVD) second-order Runge-Kutta (RK) scheme \citep[][]{SHU1988439}.
We use the dual-energy formalism \citep[DEF;][]{1995CoPhC..89..149B} to avoid numerical instabilities in the regions with highly supersonic flows.
Here we focus on the effects of magnetic amplification due to
large-scale structure (LSS) formation.
We therefore neglect radiative gas
cooling, chemical evolution, star formation, and feedback from active
galactic nuclei.
While our resolution is not suited for studying structures inside
collapsed objects, it is sufficient to capture the effects on the magnetic
cosmic web.
This is a first step in describing the evolution of more realistic PMFs than what was done previously (using the uniform seed field; see, e.g., \citealt{Marinnacietal2015,Vazzaetal2014}).
We assume a Lambda cold dark matter ($\Lambda$CDM) cosmology \citep[as in][]{Planck2018} with the
following parameters: $h=0.674$, $\Omega_m=0.315$, $\Omega_b=0.0493$,
$\Omega_{\Lambda}=0.685$, and $\sigma_8=0.807$.

%
\begin{figure}[t!]
    \centering
    \includegraphics[width=8.5cm]{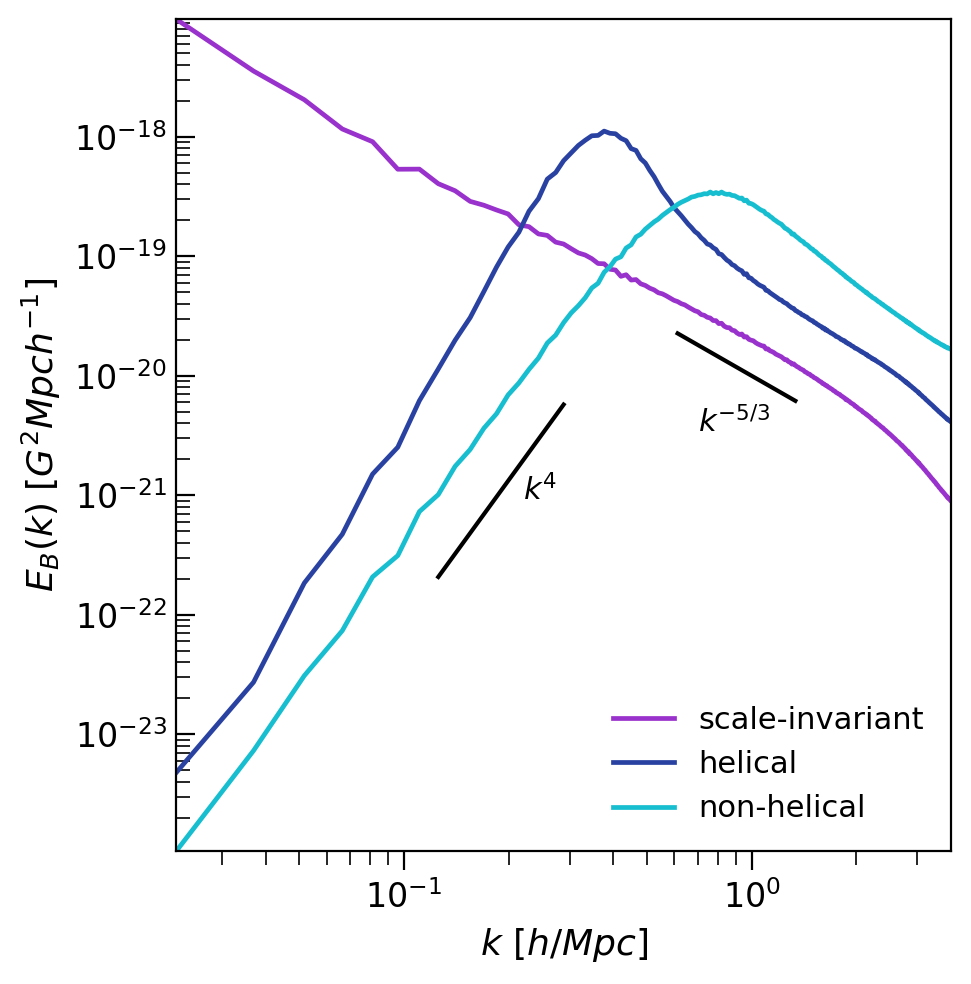}
    \caption{The initial magnetic power spectra for the stochastic setups.} 
    \label{fig:B-PS_stoch}
\end{figure}
%
%

\subsection{Initial conditions}
\label{subsec:Inits}
We consider four different scenarios for the initial magnetic seed field (see Table~\ref{tab:Tab1} and Section~\ref{sec:physMod} for the validity of the models):

\begin{enumerate}
    \item[(i)] 
    Uniform (spatially homogeneous) field: we study an initial
    seed magnetic field with a constant strength across the whole
    computational domain.
    The magnetic field is directed along the diagonal.
    This case corresponds to a particular inflationary magnetogenesis scenario, namely, the Mukohyama model \citep[][]{Mukohyama2016}.
    \item[(ii)]
     Scale-invariant field: this is a setup for a stochastic, statistically homogeneous PMF with no helicity.
     This case corresponds to an inflationary scenario.\footnote{We note that we call this model ``scale-invariant" even though it has a turbulent spectra with $k^{-5/3}$ scaling; see Section~\ref{sec:physMod}.}
    \item[(iii)]
     Nonhelical field: a stochastic, phase-transitional PMF with no helicity.
    \item[(iv)]
    Helical field: the same stochastic setup as (iii), but with helicity.
\end{enumerate}
Any power-law decay/growth of the magnetic field and its correlation
length on small scales is expected to virtually freeze at the end
of the radiation-dominated epoch \citep[][]{BanerjeeJedamzik2004},
while the large-scale evolution of PMFs during dark ages as well as
after the reionization epoch is primarily dominated by the expansion of
the universe  \citep[see][for a review]{Subramanian2016}.\footnote{See,
however, \citealt{Beraetal2020}, who claim a faster decay of PMFs
than expected by the expansion of the universe.}
We initialize the simulation at $z=50$ without loss of generality
since no significant changes at the cosmological scales of interest
for this work are expected between the recombination epoch and $z=50$
redshift.

The seed magnetic field conditions (ii)--(iv) were preproduced with the
\textsc{Pencil Code} \citep{JOSS}.
These were then used as initial conditions for the \texttt{Enzo} code.
We normalized the mean magnetic energy density in all four cases (at initial redshift $z=50$)
to the same value corresponding to a mean ({\it effective})
magnetic field strength.
The reader is referred to Appendix~\ref{AppA} for a detailed description
of the generation of the initial magnetic conditions (ii)--(iv), including the
setup for the helical field, as well as their normalization.

In Section~\ref{subsec:MFstrength}, we discuss the results of three different normalizations:
$0.1\,$nG, $0.5\,$nG, and $1\,$nG (see Table~\ref{tab:Tab1}) for the uniform,
scale-invariant, helical, and nonhelical models, respectively.
In the remainder of the paper, we only discuss the results of the $1\,$nG normalization
\citep[which is below the upper limit from the CMB bounds;][]{Plancketal2016}.

We show the initial magnetic power spectra for the stochastic setups considering a $1\,$nG
normalization in Figure~\ref{fig:B-PS_stoch}.
The initial velocities and densities are generated according to the
transfer function of \cite{EisHu1998} which accounts for the evolution of (post-inflationary) linear perturbations but neglects any effects from PMFs.
PMFs are expected to affect the
initial density power spectrum and can produce additional clustering of matter on intermediate and small scales
\citep{SethiSub2005, Yamazakietal2006, FedeliMoscardini2012,Kahniashvilietal_2013_2, Sanatietal2020}.
As was mentioned in the Introduction (Section~\ref{sec:intro}), PMFs induce gravitational instabilities,
scalar (density), vector (vorticity), and tensor (gravitational waves) modes;
for more details, see \citealt{Subramanian2016} and references therein.
In the context of the PMF's effects on LSS (or vice versa), the scalar mode plays
the most important role through modifying the matter power spectrum.\footnote{Here, the effects induced through nonlinear coupling between the modes are neglected.
We also do not consider the additional vector mode presence that is absent in the standard
cosmological scenario, and can change the initial velocity field.}
The effects on the initial matter power spectrum have been discussed in the pioneering
work of \cite{Wasserman1978}, \cite{Kimetal1996}, and \cite{GopalSethi2003}. 
All these studies have neglected PMF decay effects, assuming a frozen-in
magnetic field with unchanged spectral profile from the moment of
generation until recombination.
In our work, we consider the MHD decay effects while neglecting
the initial matter power spectrum modifications.
The self-consistent derivation of the initial linear magnetized
perturbation is beyond the scope of this paper. 
The main novelty of our approach lies in the fact that we adopt a magnetic
field from a self-consistent turbulence simulation where the field displays
an approximately self-similar decay.
In particular, the field contains coherent structures over all length scales. 

In addition, recent numerical work has
shown that such modifications to the initial power spectrum lead to effects mainly at the low end of the halo mass function \citep[see][]{Sanatietal2020,Katzetal2021}. Such effects should not significantly impact the results presented in this work since our initial magnetic conditions are better suited for studying the evolution of magnetic fields on larger scales and higher-mass haloes. Therefore, the proposed initial conditions are suitable for the questions explored in this work.

\section{Physical model}
\label{sec:physMod}
In this section, we discuss the physical motivation for studying the chosen primordial
magnetogenesis models and statistical characteristics for these scenarios.
The statistical characteristics of each PMF are determined by the generation mechanism.
First, in the case of a uniform, homogeneous PMF that is generated during the
inflationary stage, the correlation length $\lambda_B$ is undetermined (well above
the horizon scale at the generation moment). Indeed, such a PMF might be
described as a monochromatic field at $k=0$.
Such a field has zero helicity, and no spectrum can be
associated to it\footnote{It is important to emphasize that such magnetic 
fields are qualitatively different from the small-scale fields that are 
obtained by tangling of a uniform (imposed) one.
This is because, in periodic domains, such a field constitutes a separate
component that can never change.
This has dramatic consequences for the evolution of the magnetic field
on all smaller scales; see \cite{Mandaletal2020} for examples.} (i.e., 
the realization in Fourier space is a Dirac delta function).
If a statistically homogeneous field has been generated through some
mechanism after inflation, its peak scale $\lambda_{\rm peak}$ and
therefore its correlation length must be limited by the size of the Hubble
horizon at the moment of generation, $H^{-1}_\star$.
In this case, we can still use a monochromatic field at a nonzero
wavenumber $k_{\rm peak}$.
Correspondingly, the PMF in Fourier space will be approximated by 
a Dirac delta function as $\delta(k-k_{\rm peak})$.
This is a common description when referring to the lower limits 
of the magnetic field through blazar spectra observations
\citep[][]{Tayloretal2011}.
The energy density associated with a
homogeneous PMF is simply given through $B^2/(8\pi)$, regardless of whether
the field has finite or infinite correlation length.
Therefore, the initial conditions (i) and (ii)
(see Section~\ref{subsec:Inits}) represent inflationary-generated
fields with unlimited and limited correlation lengths, respectively.
In the latter case, we take into account the development
of a turbulent forward cascade in the radiation-dominated epoch
\citep[][]{Kahniashvilietal2017}.
That is, the inflation-generated, scale-invariant $k^{-1}$
spectra result in a Kolmogorov $k^{-5/3}$ spectrum by the
end of recombination.

In the case of causally generated magnetic fields after inflation,
the correlation length is limited by $H^{-1}_\star$.
Under the standard description of a PMF generated during the electroweak
phase transitions through bubble collisions, the physical correlation
length is determined by the bubble length scale.
A length scale of around one-hundredth of the Hubble horizon scale is commonly
assumed, with 100 bubbles within a linear Hubble scale.
For the QCD phase transitions, the bubble size is commonly assumed to
be one-sixth of the Hubble scale \citep[see, e.g.,][]{Schwarz2003}.
It is important to realize that the resulting decaying turbulence
is invariant under rescaling with the correlation length $\lambda_B$
such that the magnetic and kinetic energy spectra obey
\begin{equation}
E(k,t)=\lambda_B(t)^{-\beta}\phi(k\lambda_B(t)),
\end{equation}
where $\beta=1$ for nonhelical MHD turbulence and $\beta=0$ for
helical MHD turbulence \citep{BranKahn2017}.
This allows us then to rescale our initial magnetic 
field to any desired epoch based on the value of $\lambda_B$.
Therefore, at the level of the models discussed in the present paper,
there is no difference between turbulence originating
from the QCD or the electroweak phase transition.

The cases (iii) and (iv) correspond to a turbulent, causally generated
magnetic field with a characteristic scale of
$\sim 2.6 \, h^{-1}\Mpc$ and $\sim 1.26 \, h^{-1}\Mpc$, respectively.
Toward larger scales, the magnetic energy spectrum is proportional
to $k^4$, where $k$ is the wavenumber.
This spectrum is referred to as the Batchelor spectrum and occurs for
causal PMF generation \citep{Durrer+Caprini2003}.
On smaller scales, a turbulent magnetic cascade with
an energy spectrum proportional to $k^{-5/3}$ is expected.
These spectra are obtained quite generically also when turbulence is
driven monochromatically at one wavenumber \citep{Kahniashvili2020b}.
Physically significant turbulence could be generated during phase transitions
through bubble collisions (see \citealt{Subramanian2016} and \citealt{Vachaspati2020} for reviews) or through the coupling of the inflaton with the magnetic field,
possibly shortly after inflation.
%
\begin{deluxetable}{c c c }
\label{tab:Tab2}
\tablecaption{Mean of the magnetic field strength at $z=0.02$ achieved for all our models (see also Table~\ref{tab:Tab1}). Note that the mean value is computed in the whole cosmological box.}
\tablehead{
\colhead{model} & \colhead{Normalization} &  \colhead{$\langle {B\rangle_{z=0.02}}$} \\
\colhead{} &\colhead{(nG)} &\colhead{(nG)}
 }
\startdata
{}  &  $0.1$ & $0.15$ \\
Uniform  & $0.5$ & $ 0.69$  \\
{}  &  $1$ & $1.29$ \\
\hline
{}  &   $0.1$ & $0.15$ \\
Scale-invariant   &   $0.5$ & $0.69$ \\
{}  &   $1$ & $1.27$ \\
\hline
{}  &   $0.1$ & $0.09$ \\
Helical  &   $0.5$ & $0.42$ \\
{}  &   $1$ & $0.79$ \\
\hline
{}  &   $0.1$ & $0.08$ \\
Nonhelical  &   $0.5$ & $0.39$ \\
{}  &   $1$ & $0.74$  \\
\enddata
\end{deluxetable}
%

Following the results from \cite{Brandenburgetal2017},
the MHD turbulent decay of phase-transitional fields in the radiation-dominated epoch leads to an increase of the magnetic correlation length
as a function of conformal time $\eta$.
If the initial magnetic field is nonhelical, the increase is  $\lambda_B \propto \eta^{1/2}$, whereas if it is fully helical, the increase is $\lambda_B \propto \eta^{2/3}$.
This increase is expected from the moment of generation until recombination.
If a magnetic field is generated at the horizon scale with a strength limited by BBN,
the authors estimate that a fully helical (nonhelical) field may reach a (comoving)
magnetic field strength of $0.3\nG$ ($3\times10^{-3}\nG$) at a scale of $30\kpc$ ($0.3\kpc$)
at the epoch of recombination \citep[see Figure~11 in][]{Brandenburgetal2017}.
The freely decaying turbulence regime is terminated as the baryonic fluid
becomes neutral after recombination; then the comoving amplitude,
spectral shape, and helicity of the magnetic field stay unchanged until
reionization ($ z\lesssim 15$).

Our selected initial characteristic scales for models (iii) and
(iv) are larger than 
the mentioned predictions
due to our limited
resolution of 132 $h^{-1}\kpc$.
It is therefore important to stress that our initial stochastic, helical and nonhelical spectra are intended only to emulate the shape
that is expected theoretically.
In the helical case, we selected a correlation length that is larger
than the one in the nonhelical scenario since this is also expected
from theory as a consequence of the inverse cascade that helical fields
undergo in the radiation-dominated epoch \citep[see][]{Subramanian2016}.

\section{Results}
\label{sec:Results}

\subsection{Magnetic field strength}
\label{subsec:MFstrength}

We start our analysis by studying the effects of the initial magnetic field strength on the final distribution of magnetic fields in different cosmic environments. In this section, we discuss the results from three different initial magnetic strengths:
$0.1\,$nG, $0.5\,$nG, and $1\,$nG (see Table~\ref{tab:Tab1}).

In Figure~\ref{fig:B-rho-T}, we show the median of the magnetic field 
and temperature distributions with respect to the gas density distribution for 
the different models at $z=0.02$. 
In the same figure we also show a 2D histogram, i.e., we overplot
the gas mass (for the helical case) falling into each bin. The distributions differ according to the 
different regions of the cosmic web, i.e., voids, filaments, bridges (the latter
two indicated with the shaded region), and clusters.
It should be noted that the term ``bridge'' has recently been
applied to regions connecting two merging clusters.
We identified bridges in our simulations by visual inspection. 
The typical range of the overdensity in this environment is 
given in Figure~\ref{fig:B-rho-T}. 
Notably, we define bridges in a way so that they also include the outskirts of galaxy clusters.
It is important to distinguish this environment from large-scale
filaments, because currently bridges are the most promising way to observationally detect large-scale warm–hot gas filaments
\citep[e.g.,][]{Govonietal2019,2020arXiv201208491R}.

The resulting differences between magnetic seedings in each cosmic environment 
observed in Figure~\ref{fig:B-rho-T} can be summarized as follows:
\begin{itemize}
    \item[] \textit{Clusters}. As we can see from the top panel of
    Figure~\ref{fig:B-rho-T} the variation of the initial magnetic field
    strength does not affect the normalized\footnote{Normalized by the
    corresponding mean field value of the model; see Table~\ref{tab:Tab2}.}
    magnetic field distributions in galaxy clusters; although each
    model leads to higher final mean values for the higher initial
    magnetic field realization (see Table~\ref{tab:Tab2}).
    The highest median values are 
    observed from the inflationary seedings where the uniform seeding shows a magnetic field strength twice as large as the scale-invariant model. On the
    other hand, the phase-transitional seedings show the lowest magnetic field strengths within these regions. In particular, the nonhelical case is the model that shows the lowest amplification
    in galaxy clusters. 
    Interestingly, different seed magnetic fields, as well as the variation of the seed field strength, do not alter the temperature distribution in these regions
    \citep[see][for comparison]{Vazzaetal2020}.

    \item[] \textit{Bridges}. 
    Similarly to clusters, the highest magnetic fields
    in these regions are seeded by the inflationary scenarios.
    However, the differences in the magnitude of the uniform and scale-invariant cases are reduced in comparison to cluster regions, while for phase-transitional models they are slightly enlarged.
    Finally, the temperature distribution is again unaffected by
    different primordial scenarios.

    \item[] \textit{Filaments}. The different initial magnetic field strengths do not alter the normalized magnetic field or temperature trends in this region. 
    However, the final distribution of the magnetic field is affected by the initial topology of the seed field. The seed fields with initial larger characteristic scales lead to larger magnetization levels in filaments.
    The uniform and scale-invariant cases show amplification due to adiabatic contraction. 
    In the helical and nonhelical models, where the magnetic power of the large-scale modes is smaller (see Figure~\ref{fig:B-PS_stoch}), the amplification due to adiabatic compression is less efficient. 
    Previous work that focused on filaments \citep[e.g.,][]{Vazzaetal2014}, 
    have argued that the magnetic amplification within these regions is less dependent on resolution and possible dynamo action even at high Reynolds numbers. This can happen because compressive modes are expected to be dominant in filaments and suppress small-scale twisting of magnetic structures.
    It remains interesting for future work to assess whether our helical and nonhelical models
    at higher resolution and, therefore, higher Reynolds number could further amplify magnetic fields by promoting more solenoidal modes within filaments.
    Finally, it seems that the differences between the inflationary and phase-transitional models are not reflected in the temperature distribution. Similar trends have been found in \cite{GhellerVazza2019}, where the different seeding scenarios show significant imprints on final magnetic field distributions but only mild differences in the temperature trends (pronounced mostly at high densities).
\begin{figure}[htbp]
    \centering
    \includegraphics[width=8.6cm]{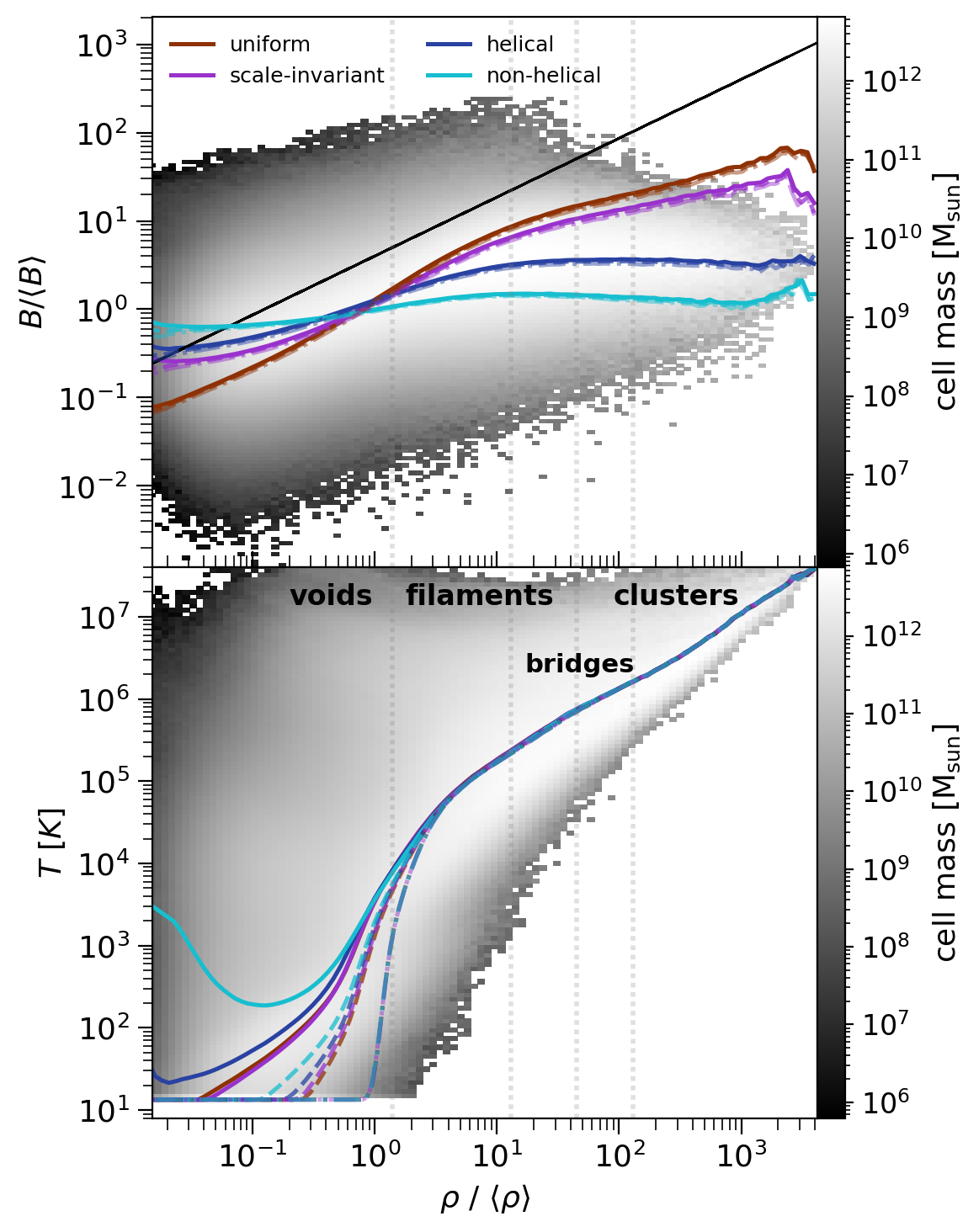} 
    \caption{
    Dependence of the median magnetic field and temperature on density
    for all of our simulations.
    The $x$-axis shows the gas density normalized by the mean density field.
    The solid, dashed, and dashed-dotted lines correspond to the $1$ nG, $0.5$ nG, and $0.1$ nG,
    normalizations, respectively.
    The black solid lines show the expected density scaling of the
    magnetic field strength 
    based on the adiabatic contraction only
    ($\propto \rho^{2/3}$). 
    The additional color coding (black–white palette) shows the mass of gas (for the helical case) falling into each bin. 
    Vertical dotted lines indicate the characteristic densities in filaments
    and bridges.
} 
    \label{fig:B-rho-T}
\end{figure}
%
    \item[] \textit{Voids}. 
    These regions are assumed to be most promising for discriminating among different magnetogenesis scenarios \citep[see, e.g.,][]{DubTey2008,Donnertetal2009}. 
    We see that differences between the models in the temperature profiles are revealed only for the higher initial magnetic field realizations (i.e., only for the $0.5\,$nG and 
    $1\,$nG initial magnetic field strengths). 
    However, the differences between the uniform and scale-invariant
    models are negligible, while the helical and nonhelical cases show the largest discrepancies.
    In addition, we see that stochastic (scale-invariant, helical, or nonhelical) seedings lead to the highest magnetic field strengths, and consequently highest temperatures within these regions. 
    This highlights the fact that initial 
    stochastic turbulent motions could cause local heating. 
    Similar trends were obtained by \cite{Vazzaetal2020} where only models with an initial power-law magnetic spectrum were considered. Therefore, initial magnetic fluctuations could be important for the local heating dynamics. In particular, PMFs can heat up the hydrogen and helium in the IGM by, for example, decaying turbulence \citep[see, e.g.,][]{SethiSub2005,Minodaetal2017} and this can be reflected in the 
    \HI{} 21 cm signal \citep[][]{Sethi2005,Minodaetal2019,Natwariya2021}. Strong absorption spectra of the global \HI{} signal detected by the EDGES experiment suggest that IGM was colder than expected from the standard cosmological scenario \citep[][]{Bowmanetal2018}. 
    In the recent work of \cite{Beraetal2020} it has been shown that heating of IGM due to PMFs along with DM–baryon interactions \citep[][]{Barkana2018} can be used for constraining the strength of PMFs.
    Hence, the observed temperature differences in Figure~\ref{fig:B-rho-T} could be used to probe the strengths of both inflationary as well as phase-transitional models in cosmic voids.
\end{itemize}

\begin{figure*}[htbp]
    \centering
    \includegraphics[width=18.6cm]{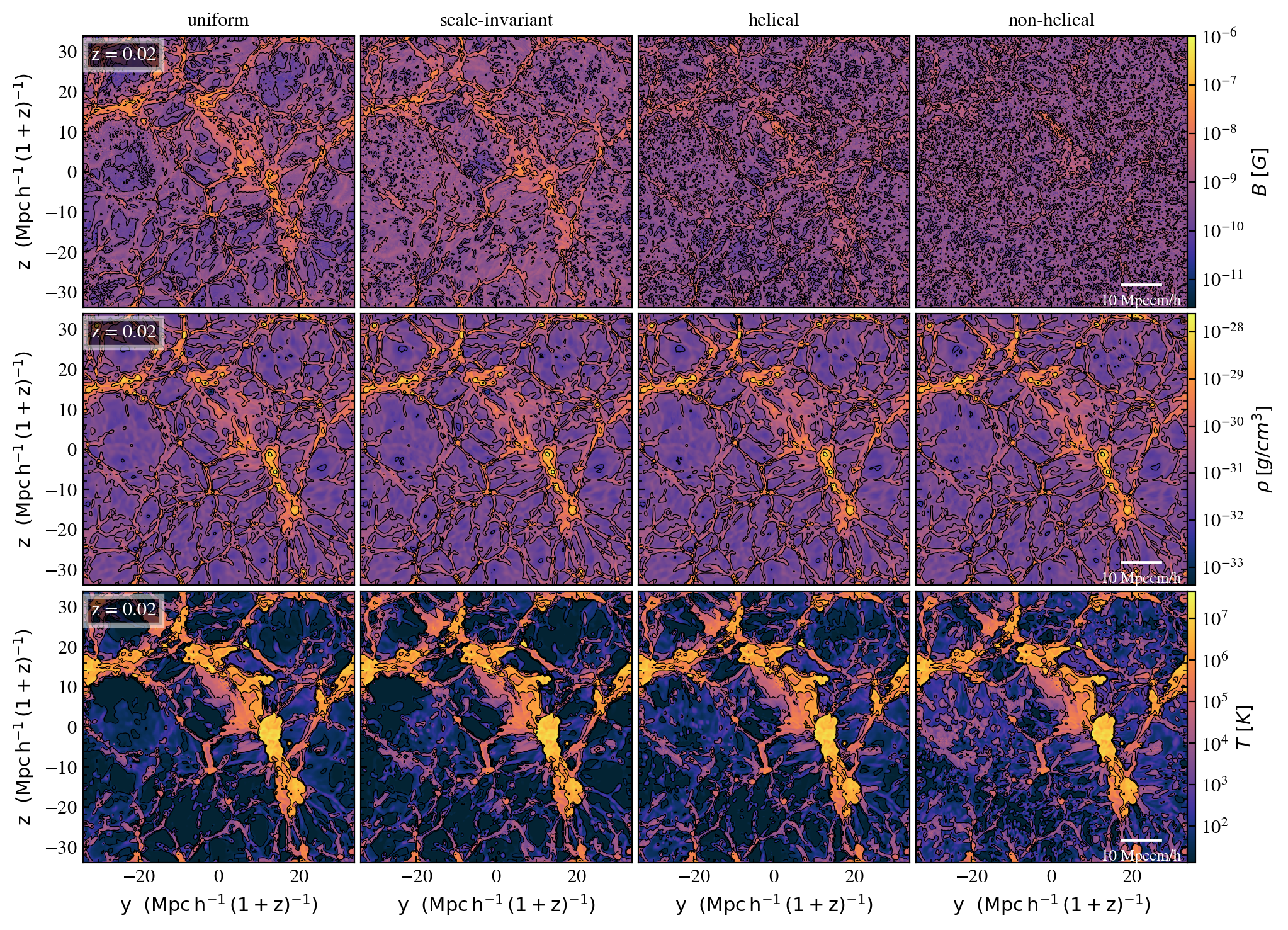}
    \caption{Contoured slices through the
    center of the simulated box at $z=0.02$.
    The top, middle, and bottom panels show the magnetic field, density,
    and temperature slices correspondingly.
    The overplotted contour lines mark the regions with a certain field
    strength, and the range of the field values are set according to the
    minimum and maximum of the annotated fields.}
    \label{fig:Proj2}
\end{figure*}
We can conclude that the amplification of the
initial magnetic field due to adiabatic contraction is 
subdominant in the stochastic turbulent cases at our resolution. Conversely, the uniform model
can be well fitted by the simple $\rho^{2/3}$ relation (expected from adiabatic contraction only) in the filaments and low-density regions (see also Appendix~\ref{App:ResTest}). 
As expected, the differences between models become larger at the galaxy-cluster scales due to our limited resolution. While we expect that these differences should persist (although to a lesser degree) with higher resolution, only future work focusing on these regions can demonstrate it.

For the remaining part of the paper we will focus 
on the PMF scenarios with an initial $1\,$nG normalization (see Table~\ref{tab:Tab1}).
The resulting mean magnetic field strengths from this normalization are more in agreement with current observations; see, e.g., \cite{FuscoFemiano2001,FuscoFemiano2004} for the
$0.1$--$1\,\mu$G magnetic field strengths found in galaxy clusters 
based on the inverse Compton measurements from radio haloes, and
\cite{Vogt2005} for the strengths as high as $1$--$7\,\mu$G based on the
likelihood analysis of Faraday rotation measures.

\subsection{General properties}
\label{sec:general}
We show slices of the final ($z=0.02$) magnetic field,
density, and temperature in the
$67.7 \, h^{-1}\Mpc$ box in Figure~\ref{fig:Proj2}.
As we can observe in the top panels of Figure~\ref{fig:Proj2}, the
stochastic seeding fills the voids more efficiently than the uniform
case, while the uniform and scale-invariant magnetic seedings show the
largest spatial correlation with the filamentary structures.
This is not surprising because the initial magnetic power spectrum
of the uniform seeding does not have a characteristic scale, and the
characteristic scale of the scale-invariant spectrum is larger than that of the helical and nonhelical cases.
We also observe that nonhelical initial magnetic fields
tend to create more substructures than helical ones.
This is an expected trend in MHD simulations of helically
and nonhelically driven turbulence because
of the magnetic helicity (and thereby magnetic energy) transfer to large scales.
In a recent study of turbulent dynamos driven by isotropic forcing in isothermal MHD \citep{Vaisalaetal2020}, helical and nonhelical cases were compared. The authors found that an initially Gaussian magnetic field
(with a magnetic energy spectrum $\propto k^2$) develops a magnetic
energy spectrum with a characteristic scale (peak spectrum) in both
cases. Nonetheless, the helical case attains a large-scale nature (more
power at $k$ smaller than the peak scale), while the nonhelical case
is still characterized by a small-scale structure.

The density (middle row in Figure~\ref{fig:Proj2}) and temperature (bottom row in Figure~\ref{fig:Proj2}) contours show a spatial correspondence between
the densest structures ($\sim10^{-28}\g\cm^{-3}$) and the
highest temperatures ($\sim10^{7} \text{K}$) in all four cases.
The low-density regions ($\sim10^{-32}\text{--}10^{-31}\g\cm^{-3}$)
enclose higher temperatures in the nonhelical case than
in all the other cases.
This is a direct consequence of having smaller magnetic substructures in the
nonhelical model at the initial and final redshifts, which leads to extra turbulent dissipation in the voids region.
\begin{figure}[t!]
    \centering
    \includegraphics[width=8.36cm]{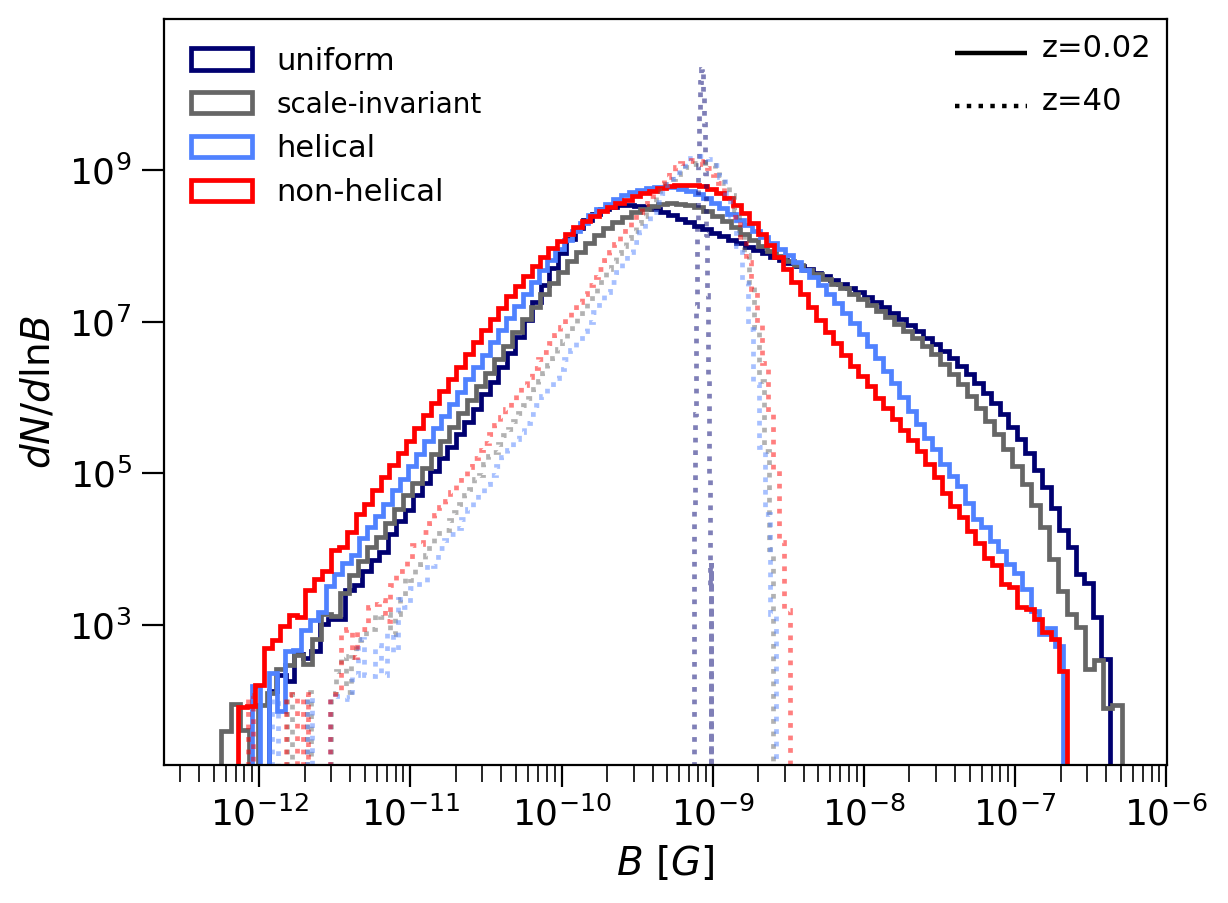}
    \caption{Probability distribution functions for the different magnetic seedings at $z=40$ (dotted lines) and $z=0.02$ (solid lines).}
    \label{fig:PDFs}
\end{figure}

In Figure~\ref{fig:PDFs} we show the volumetric probability distribution
functions (PDFs) of the magnetic field for all our models. 
The PDFs are shown for two epochs: $z=40$ (dotted lines) and $z=0.02$
(solid lines).
The final PDFs show a broadening in all the models.
The low-end tail of the distribution (values below $6\times 10^{-10}\G$) is
very similar for all the models.
By contrast, the high-end tail of the distribution (values above $4\times 10^{-9}\G$) differs in the inflationary and phase-transitional models.
Both inflationary models produce higher magnetic field values than the
phase-transitional models.
It should also be noted that the PDFs for both scenarios deviate
from the Gaussian trend (more evidently seen for the inflationary models) and the peak of the distribution in the inflationary case is shifted
toward lower values.
For example, in the uniform case, the peak is shifted from
$8\times 10^{-10}\G$ to $2\times 10^{-10}\G$.
We can see that the main differences in the distributions come from the
regions where the magnetic field strength is of the order of
$4\times 10^{-9}$--$5\times 10^{-7}\G$.

\begin{figure}[htbp]
    \centering
    \includegraphics[width=8.4cm]{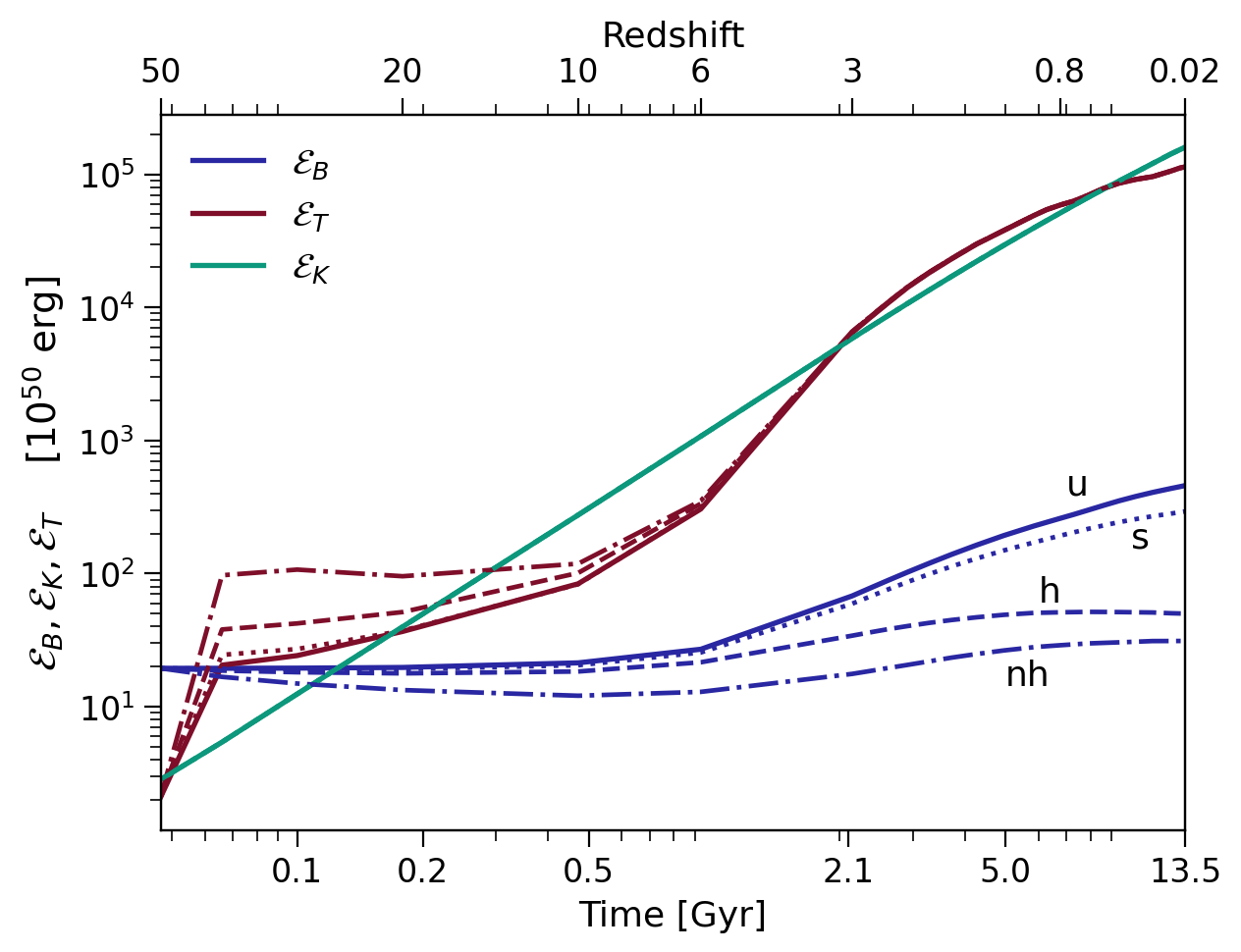}
    \caption{Evolution of magnetic, thermal, and kinetic energies for the different magnetic seedings. The solid, dotted, dashed, and dashed-dotted lines correspond to uniform, scale-invariant, helical, and nonhelical cases, respectively.}
    \label{fig:EnergyEv}
\end{figure}

Finally, the evolution of thermal, kinetic, and magnetic energies
over a time span of 13.5 Gyr is shown in Figure~\ref{fig:EnergyEv}.
The thermal $\mathcal{E}_T$, kinetic $\mathcal{E}_K$, and magnetic $\mathcal{E}_B$ energies are defined as 
%
%
\begin{equation}
    \int n k_B T dV, ~~~ 
    \int \frac{\rho {\mathbf{v}^{2}}}{2} dV, ~~~ 
    \int \frac{\mathbf{B}^2}{8\pi} dV,
\end{equation}
where $n$ is the gas number density, $\mathbf{v}$ the velocity, $k_B$ the Boltzmann constant, and $V$ the volume.
The growth of kinetic and thermal energies achieved in all four models
is of the order of $\sim 10^5$ by the end of the simulation.
The thermal energies show variations between the models at earlier redshifts
(see $z\gtrsim 6$, where stochastic phase-transitional cases show larger thermal energies).
This indicates that more dissipative processes are present in these scenarios.
Although we find that the extra heating shown in Figure 5 is independent of the Riemann solver, we cannot rule out a numerical origin. However, we also see that the thermal-energy evolution converges to the hydrodynamic case for redshifts $z\lesssim 6$ (see also Appendix~ \ref{App:RSTest}). For this reason, it is unlikely that our analysis at lower redshifts is affected by the initial transient. Moreover, spurious heating will be suppressed by radiative cooling in a more advanced model.

\begin{figure*}[htbp]
    \centering
    \includegraphics[width=18.0cm]{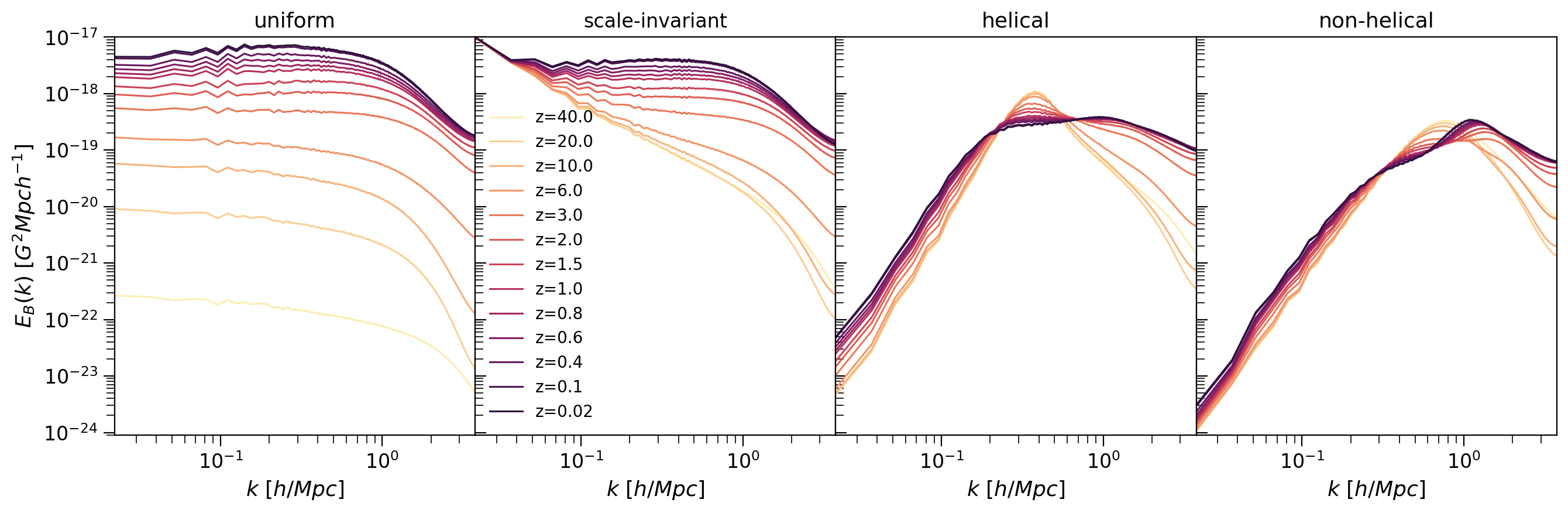}
    \caption{Redshift evolution of magnetic power spectra. From left to right: the uniform, scale-invariant, helical, and nonhelical seedings. } 
    \label{fig:B-PS}
\end{figure*}

The final thermal and kinetic energies (for $z\gtrsim 6$) show no difference
between the four models.
However, differences in the magnetic energies arise at a redshift as early as $z\sim40$.
At $z=0.02$ it is evident that the magnetic energy in the inflationary models is roughly 
one order of magnitude larger than the magnetic energy in the
phase-transitional models.
Overall, the magnetic energy growth throughout the evolution is observed to
be by a factor of $\sim 10$ in the uniform and scale-invariant cases, and by
a factor of $\sim 2$--$4$ in the nonhelical and helical cases, respectively.

\subsection{Evolution of magnetic power spectra} 
\label{subsec:MagPS}

In this section, we focus on the properties of the magnetic power spectra for the different primordial magnetic seedings.
We define the magnetic energy power spectrum in Fourier space as
\begin{equation} \label{E_B}
\int E_B(k) dk=\frac{1}{2V}\int \hat{\textbf{B}} \cdot \hat{\textbf{B}}^* 4\pi k^2 dk,
\end{equation}
where $\hat{\textbf{B}}$ denotes the Fourier transform of the magnetic field,
$\hat{\textbf{B}}^*$ is its complex conjugate, $k=|\textbf{k}|$ is the norm of
the wavevector and $V$ is the volume that normalizes the spectrum.

In Figure~\ref{fig:B-PS} we show the evolution of the magnetic power spectra
of the inflationary (uniform and scale-invariant) and
phase-transitional (helical and nonhelical) cases.
Notable features of this figure are as follows:
\begin{itemize}
    \item[a)] \textit{Inflationary seeding}. The magnetic power spectrum
    in the uniform and scale-invariant cases reaches higher values
    at the final redshift than that of the cases with
    helical and nonhelical seeding.
    In particular, the uniform seeding shows how the magnetic energy builds up at all
    scales in a similar fashion from early redshift ($z=40$).
    This emphasizes how the magnetic field follows the growth of the
    density perturbations. 
    On the other hand, the amplitude of the power spectrum in the
    scale-invariant case starts increasing only from $z=10$,
    but the final distribution achieved in this case is similar
    to that in the case of the uniform seeding.
    
    The obtained trends for these two subcases of an inflationary
    scenario seem to be different from the results of a recent study
    by \cite{Mandaletal2020}, where the MHD evolution with
    imposed and scale-invariant initial fields has been
    compared in the radiation-dominated epoch.
    They showed that even if small-scale turbulent forcing
    is applied, the uniform (imposed) field always decays
    faster than the field with the scale-invariant spectrum.
Subsequent studies revealed that the apparent difference
between the two types of simulations is caused by the fact that
in the present cosmological simulations there is always a
large-scale velocity field, which was not the case in the
simulations of \cite{Mandaletal2020}.
However, repeating their simulations with a large-scale velocity
field characterized by an initial $k^{-2}$ spectrum produces the 
rapid growth of the magnetic field also on large scales; see
Appendix~\ref{App:Tangling}, where we demonstrate the tangling of a
homogeneous magnetic field by an initial turbulent velocity field with
a $k^{-2}$ spectrum.
Thus, there is no conflict between these two types of simulations
if comparable initial velocity fields are used in both cases.
    \item[b)] \textit{Phase-transitional seeding}. The initial
    characteristic scale $\lambda_{\rm peak}$ in these cases makes the magnetic power
    spectra evolve in a very different way.
    Overall, the total amplitude is smaller than in the inflationary cases, as can be observed in Figure~\ref{fig:B-PS}. 
    Nevertheless, the most interesting result is how the characteristic
    scale defines the evolution at large and small scales.
    On large scales ($k \lesssim 1\,h\Mpc^{-1}$),
    the magnetic field growth is moderate.
    The helical seeding shows stronger magnetic growth than the
    nonhelical seeding.
    This happens because in the former case the initial magnetic
    perturbations are correlated on larger scales
    (see Figure~\ref{fig:B-PS_stoch}).
    It is also possible that helicity leads to
    larger power on these scales.
    This is an expected trend in MHD simulations of decaying turbulence
    where the larger growth is observed for large length scales
    due to the inverse transfer from small to larger scales; see,
    e.g., \cite{BanerjeeJedamzik2004} and \cite{Brandenburgetal2015}.
    On the other hand, scales smaller than the characteristic scale, i.e.,
    $k\gtrsim 0.4\,h\Mpc^{-1}$ for the helical case and $k\gtrsim 0.9\,h\Mpc^{-1}$
    for the nonhelical case, respectively, can grow and reach magnetic
    levels comparable to the inflationary cases.
    Additionally, we observe that there is a shift of the peak spectra
    toward smaller scales (at $ z\sim 5$) in both cases as more substructure is building
    up on galaxy-cluster scales ($\sim 1\Mpc$).\footnote{It should be noted that after the first shift of the peak,
    the peak again gradually moves from smaller to larger scales
    (from $z=3$); this behavior is similar to the aforementioned inverse
    cascade. This trend is more evident for the nonhelical case.}
    Finally, we also observe magnetic power decay 
     on the peak scales and on the smallest scales at early redshifts ($z>10$) in both the helical and nonhelical cases. 
     In these cases, most of the magnetic energy is initially contained toward small scales while the matter (and density) power spectrum has most of its power contained on larger scales.
     This limits, for example, the magnetic energy growth in the two cases at early redshifts (see also the discussion at the end of this section). 
\end{itemize}

\begin{figure}[htbp]
    \centering
    \includegraphics[width=8.5cm]{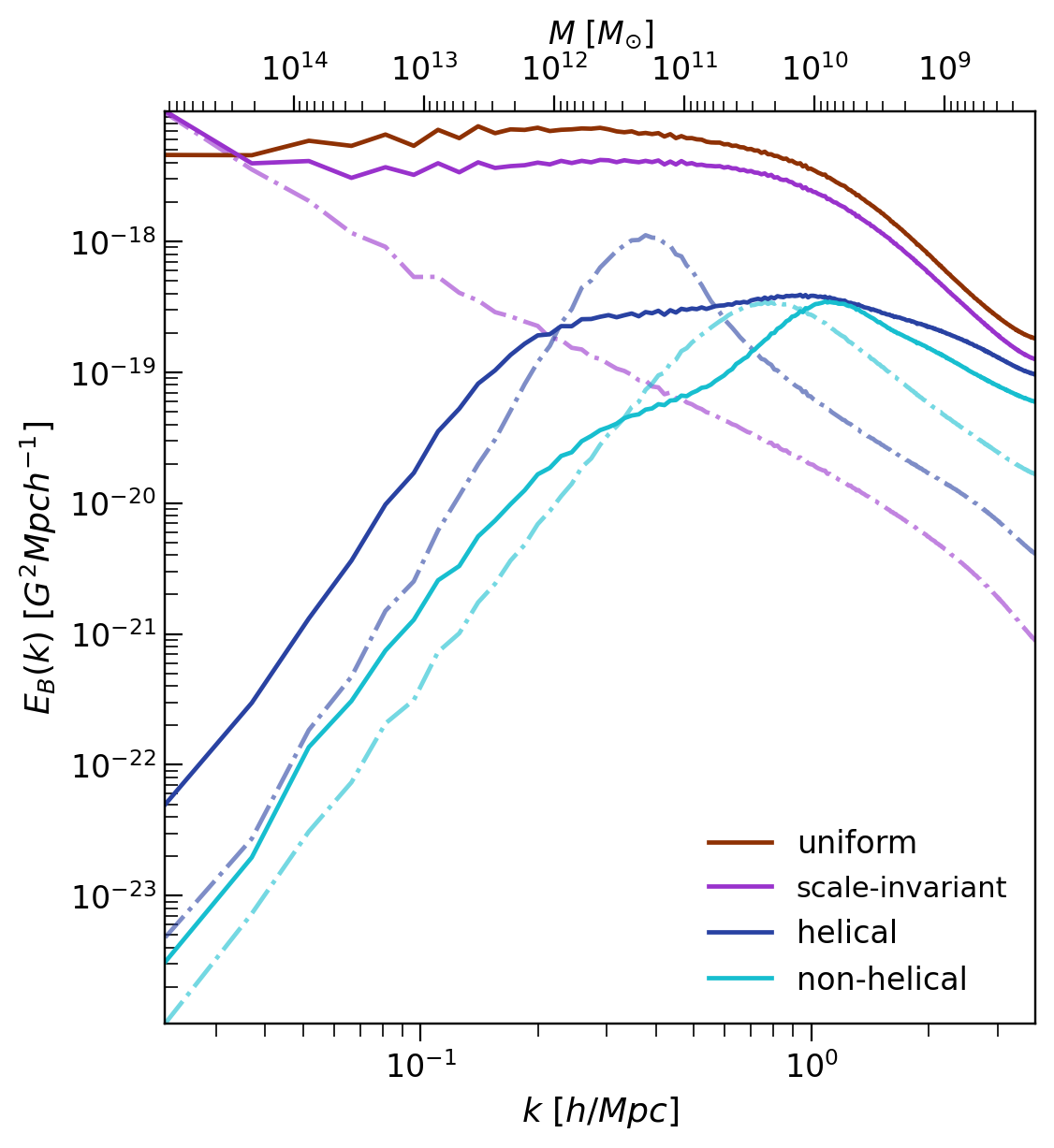}
    \caption{Magnetic power spectra for
    the uniform and stochastic cases. The dashed-dotted lines show the
    corresponding power spectra at the initial redshift ($z=50$) and
    the solid lines at the final, $z=0$ redshift.}
    \label{fig:B-PS-compM}
\end{figure}
We compare the initial and final magnetic power spectra
with the corresponding linear mass scales for all the models in
Figure~\ref{fig:B-PS-compM}.
Note that these mass scales should not be interpreted as the masses
of the massive objects (e.g., galaxy clusters) since they can only be
accepted as an estimation of masses at certain radii that are still in
the linear regime
(i.e., these would be the largest scales in our simulations, 
$k \lesssim 0.5\,h\Mpc^{-1}$).
The difference between the amplitudes of the inflationary and 
phase-transitional magnetic power spectra is more evident at small wavenumbers. 
As we mentioned above, in Figure~\ref{fig:B-PS-compM} the spectra peak of the two stochastic seedings are shifted toward larger
wavenumbers. In addition, we see that different phase-transitional seedings are expected
to be harder to distinguish at masses $M \lesssim 10^{10} \, M_{\odot}$,
whereas the magnetization at masses $M \gtrsim 10^{12} \, M_{\odot}$ is
distinguishable for the inflationary and phase-transitional models.
The reader may note that the behavior of the magnetic amplification at different scales is strongly affected by the spectrum peak
(or by the coherence scale)
at the initial redshift.
The role of the peak position will be considered in future work.

The differences between the uniform and scale-invariant spectra for all mass scales is remarkably small. It is only at 
$3 \times 10^{14} \, M_{\odot}$ that the amplitude of the scale-invariant spectra is higher than that of the uniform model.
Overall, we discern these differences in the amplitude of the power spectrum between the inflationary and phase-transition models to be increased on larger scales ($ M \gtrsim 2 \times 10^{10} \, M_{\odot}$) reaching order of $10^5$ difference on mass scales $ \gtrsim 10^{14} \, M_{\odot}$.

In a collapsing magnetized region, field amplification mainly
occurs via adiabatic contraction. However, if that region contains a
randomly oriented magnetic field, there are two additional things to consider: 
1) the cancellation of opposite-polarity fields can reduce the magnetic
flux, and/or 2) there could be extra field amplification by turbulent
dynamo if the growth rate is faster than the gravitational compression
rate (see also \citealt{SetaFed2020} and Appendix A in \citealt{Setaetal2018} for a comparison between uniform and stochastic models in idealized MHD simulations).
In our study, shocks that originate during structure formation can
additionally affect the magnetic amplification since they can destroy
coherence small-scale structures. 
This can also contribute to decreasing power of the peak scales of helical and nonhelical cases at early redshifts.
All these effects explain why in the helical
and nonhelical cases we see less efficient amplification on both small
and large scales (see Figure~\ref{fig:B-PS}). 
While all our four cases are affected by the not-well-resolved turbulent motions within the collapsing regions of the cosmic web, the lower magnetic power increase observed in the two phase-transitional cases 
could be partly
attributed to field cancellation (see also discussion below on the correlation length).
On the other hand, the stochastic, scale-invariant case develops a nonzero mean field 
(due to an initial larger correlation length in this case)
which makes its
evolution very similar to that of the uniform case and less subject to
field-cancellation effects. Hence, we observe a larger magnetic field amplification
in this case.

\begin{figure}[t!]
    \centering
    \includegraphics[width=8.5cm]{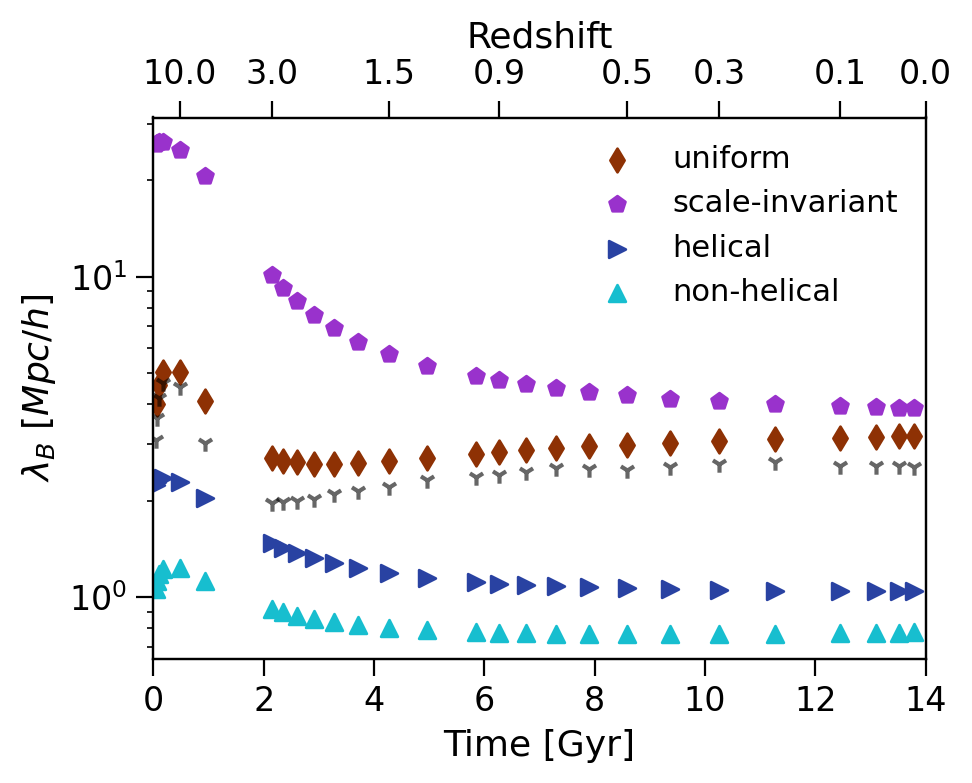}
    \caption{Evolution of magnetic correlation length obtained from different magnetic seeding. Gray points show the correlation length computed from the density power spectrum.} 
    \label{fig:CorLength}
\end{figure}

Finally, we show in Figure~\ref{fig:CorLength} the evolution
of the magnetic correlation length, defined as
\begin{equation}
    \lambda_B = \frac{\int_0^\infty dk\, k^{-1} E_B(k,t)}{ \int_0^\infty dk \, E_B(k,t)}.
\label{CorLength}
\end{equation}

It should also be emphasized that for inflationary, uniform magnetic
fields, the correlation length at the initial redshift can be ill-defined,
because the numerator of Equation~(\ref{CorLength}) diverges for $k\to0$,
so this point needs to be excluded.
For our nearly scale-invariant field, however, the spectral energy
goes to zero for $k\to0$, and thus the integral does not diverge; see
\cite{Brandenburgetal2018} for details.

In all four models, we see an increasing trend at initial redshifts which
mostly follows the evolution of the density correlation length.
In unigrid Eulerian cosmological simulations, the density power spectrum
tends to be more damped at small scales since the gravity forces are
smoothed at the grid scale, as noted by \citealt{HahnAbel2011}.
This slower growth of the smallest scales leads to the increase
of the magnetic and density correlation length at $z >10$ in
Figure~\ref{fig:CorLength}.

The correlation-length evolution is followed by a decrease
($z \lesssim 10$) in both density and magnetic correlation lengths for
all four models.
This results from the small-scale modes entering the nonlinear
regime of the growth of density perturbations.
At this stage, power is transferred from large to small scales.
A final increase of magnetic correlation length is only noticeable in
the uniform seeding case at $z \lesssim 1.75$.
The trends in the redshift range $1.75-0$ follow $t^{0.16}$,
$t^{-0.34}, t^{-0.11}$, and $t^{-0.03}$ for the uniform, scale-invariant,
helical, and nonhelical cases, respectively.
The density correlation length shows a similar trend as the magnetic
correlation length in the uniform case, although with a slower increase
with $t^{0.13}$.
It should be noted that the latter trend could also be affected by
the damped growth of perturbations at small scales because of our limited resolution.

Finally, we see that the correlation length in the uniform case is
about twice as large as in the helical and nonhelical cases, and
the scale-invariant model shows the largest final correlation length
($\sim 4\,h^{-1}\Mpc$).
This is in line with our previous discussion on the discrepancies in
magnetic amplification of inflationary and phase-transitional cases.

 \subsection{Faraday rotation measures}
 \label{sec:RM_analysis}
Faraday rotation of linearly polarized sources is a powerful
observational approach that can help us understand magnetic
fields on a vast range of astrophysical scales.
Polarized radio emission from distant sources is affected by an
intervening magnetized plasma which rotates the intrinsic
polarization plane. The differential angle of rotation
is proportional to the square of the
emitted wavelength $\lambda^2$ and to the rotation measure (RM). For a source at cosmological distance and redshift $z$, RM is defined as
\begin{equation}
\begin{aligned} 
\label{eq:RM} 
\mathrm{RM} &=\frac{e^3}{2\pi m_\text{e}^2c^4}\int_0^{l_{s}}  (1+z)^{-2}n_e(z)B_{||}(z)dl(z) &&\\\
&= 0.812 \int_{0}^{l_{s}} (1+z)^{-2} \bigg( \frac{n_e}{\text{cm}^{-3}} \bigg)\bigg( \frac{B_l}{\mu \text{G}} \bigg) \bigg( \frac{dl}{\text{pc}} \bigg) ~~\frac{\text{rad}}{{\text{m}}^{2}}, &&
\end{aligned}
\end{equation}
where $e$, $m_e$, and $c$ are the electron charge, electron mass, and
speed of light, respectively; $n_e$ is the electron number density, $B_\|$ is the magnetic field component along the line of sight
(LOS), and $dl$ is the integration path length. In this section, we use physical quantities, such as
physical magnetic field strength, electron number density, and physical length scales. 
A positive (negative) RM implies a magnetic field pointing towards
(away from) the observer.
Note that, in general, Equation~\eqref{eq:RM} defines the Faraday depth
$\phi$, which is inferred from observations through the method of RM
synthesis \citep[][]{Burn1966,BruynBrentjen2005_1}.
The RM coincides with Faraday depth in the ideal case when the rotation
is caused by only one, nonemitting Faraday screen.

RM observations of extragalactic radio sources are
fundamental in further constraining the properties of large-scale
magnetic fields in the near future.
In particular, they will be of greatest importance for discriminating between magnetogenesis scenarios because of their possibility to constrain the magnetization in the still poorly constrained regions of the cosmic web, i.e., filaments and voids. 
As the extragalactic polarized emission travels through the magnetized medium between source and observer, there are various contributions along the LOS that add up to the total observed RM. 
One relevant contribution comes from the Milky Way
\citep[see][]{Hutschenreuteretal2021}, but there have been
recent efforts in constraining the extragalactic contribution alone
\citep[e.g.,][]{Schnitzeler2010,Oppermannetal2015,Vernstrometal2019,
OSullivanetal2020}.
These recent RM observations have constrained the extragalactic
RM variance to $~5$--$10\,$rad$\,$m$^{-2}$ at 1.4 GHz.
At lower frequencies, \cite{OSullivanetal2020} limit the extragalactic
contribution to $<1.9$ rad $m^{-2}$ (measured at 144 MHz).

\begin{figure}[htbp]
    \centering
    \includegraphics[width=8.10cm]{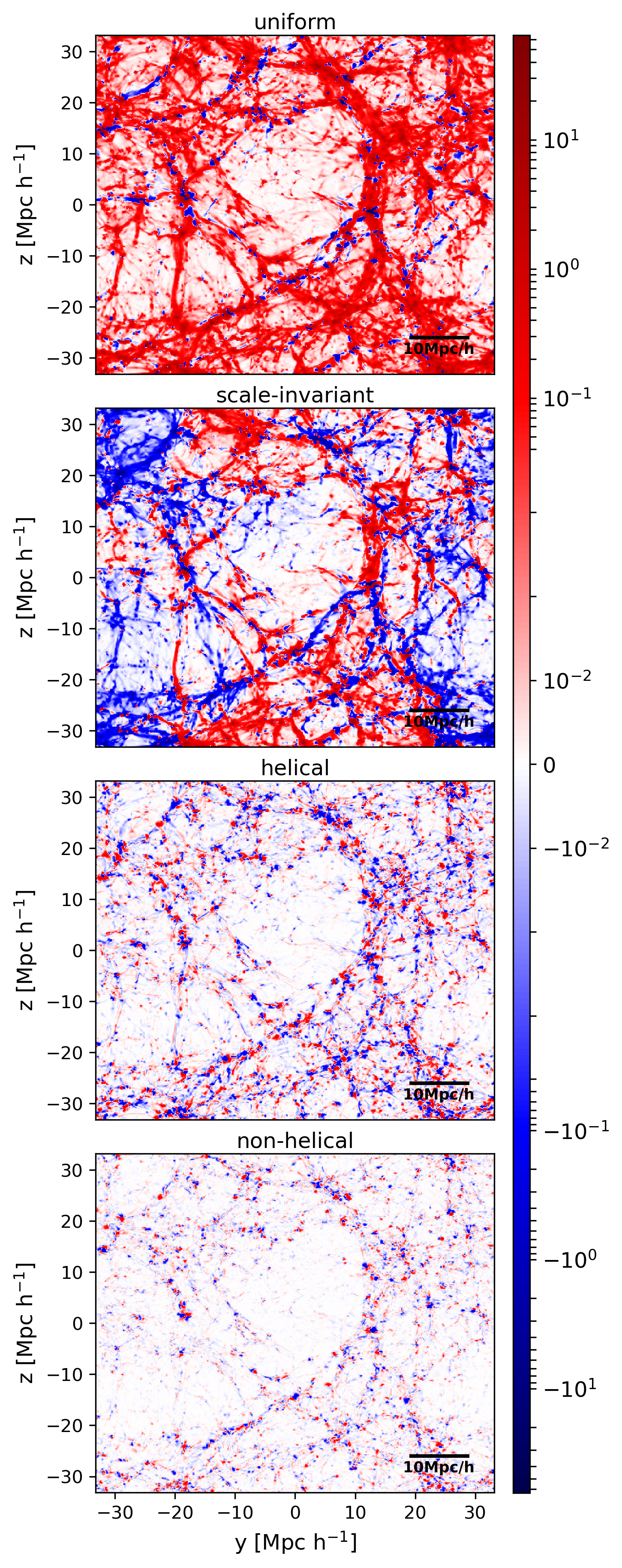}
    \caption{Faraday rotation maps from the simulated cosmic web at $z=0.02$. 
    From top to bottom: uniform, scale-invariant, helical, and nonhelical cases respectively. The color bar shows values in [rad/m$^2$] and it is linearly scaled in the range $[-0.04,0,04]$.}
    \label{fig:RM}
\end{figure}
In Figure~\ref{fig:RM} we show the RM maps of the simulated cosmic web at
$z=0.02$ for different primordial seeding cases.
Note that we did not include the Galactic contribution
\citep[][]{Hutschenreuteretal2021}. 
These maps have been obtained by integrating Equation~(\ref{eq:RM})
along the $x$-axis.
We checked that the selected axis of projection does not produce
differences in the results described in this section.
We observe significant differences in the RM maps for the different
primordial seeding models.
First, we observe more coherent structures in the uniform and
scale-invariant seeding cases than in the helical and nonhelical cases.
Second, the RM values are highest for the inflationary scenarios and
lowest in the helical and nonhelical ones.
This result is a consequence of the RM being an integrated quantity.
The sum of coherent magnetic fields will give rise to a strong
RM signal, which is the case in the uniform and scale-invariant models
(see first and second panels of Figure~\ref{fig:RM}).
On the other hand, the sum of stochastic magnetic fields can
cancel out and weaken the RM signal (see third and fourth panels
of Figure~\ref{fig:RM}).
In addition, the RM maps are determined by the total level
of magnetization at this epoch (see Table~\ref{tab:Tab2}).
The helical and nonhelical seedings lead to lower magnetization
levels in filaments, as was discussed in Sections~\ref{sec:general}
and \ref{subsec:MagPS}.
Therefore, this adds to the discrepancy observed between inflationary
and phase-transitional models.
We also note that uniform seeding leads to higher
Faraday rotation than the 
scale-invariant seeding.
This seems to be in agreement with \cite{Vazzaetal2020},
where the authors explored also a uniform seed field and various
seeds described by power laws.
Nevertheless, our inflationary results cannot be directly compared
to this recent work since they applied a subgrid dynamo model
for further magnetic amplification.

\begin{figure}[t!]
    \centering
    \includegraphics[width=8.5cm]{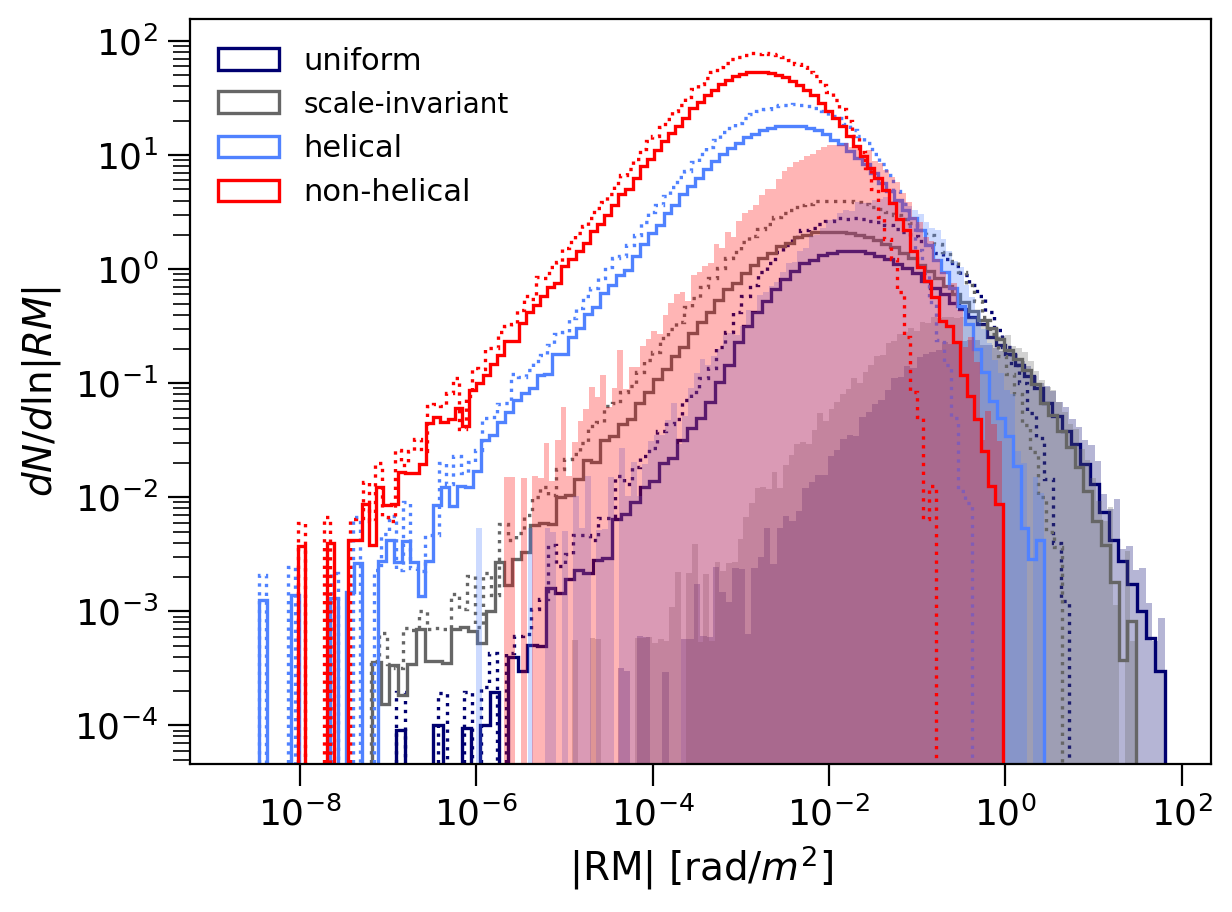}
    \caption{Distribution functions of the absolute RM for different
    seeding models. The dotted lines show the contribution to the total
    PDF from the regions excluding galaxy clusters while the
    filled histograms represent contributions from galaxy clusters
    ($\rho /\langle \rho \rangle \geq 1.3\times 10^{2}$).
    } 
    \label{fig:RM_PDF}
\end{figure}

In Figure~\ref{fig:RM_PDF} we show the corresponding PDFs
for the absolute value of RM at $z=0.02$.
We additionally show the PDF for the regions excluding galaxy clusters (dotted lines), and for galaxy clusters with the overdensity criterion $\rho /\langle \rho \rangle \geq 1.3\times 10^{2}$ (filled histograms).
These criteria were applied before computing the integral
defined in Equation~(\ref{eq:RM}).
The distributions accounting for the whole $67.7\,(h^{-1}\Mpc)^3$
region, as well as distributions excluding galaxy clusters, peak at $1.6\times 10^{-3}$, $ 4 \times 10^{-3}$, $1.2 \times
10^{-2}$, and $1.7 \times 10^{-2}$ rad m$^{-2}$ for the nonhelical, helical,
scale-invariant, and uniform cases, respectively.
When considering only the highly ionized regions (clusters; $T>10^{6}\,$K),
the PDFs peak at 
$1.5\times 10^{-2}$, $ 4.5 \times 10^{-2}$, $0.3$, and $0.5$ 
rad/m$^2$ for the nonhelical, helical,
scale-invariant, and uniform cases, respectively.
We see that the highest RM values are obtained at these highly ionized regions for the four models. 
As can also be seen in Figure~\ref{fig:RM}, the highest
values of RM tend to follow the collapsed structures.
Similar trends have been observed in other cosmological simulations
\citep[see, e.g.,][]{Marinnacietal2015}.
In line with Figure~\ref{fig:RM}, we find that the highest RM values
are observed for the two inflationary models.

\begin{figure*}[htbp]
    \centering
    \includegraphics[width=6.1cm]{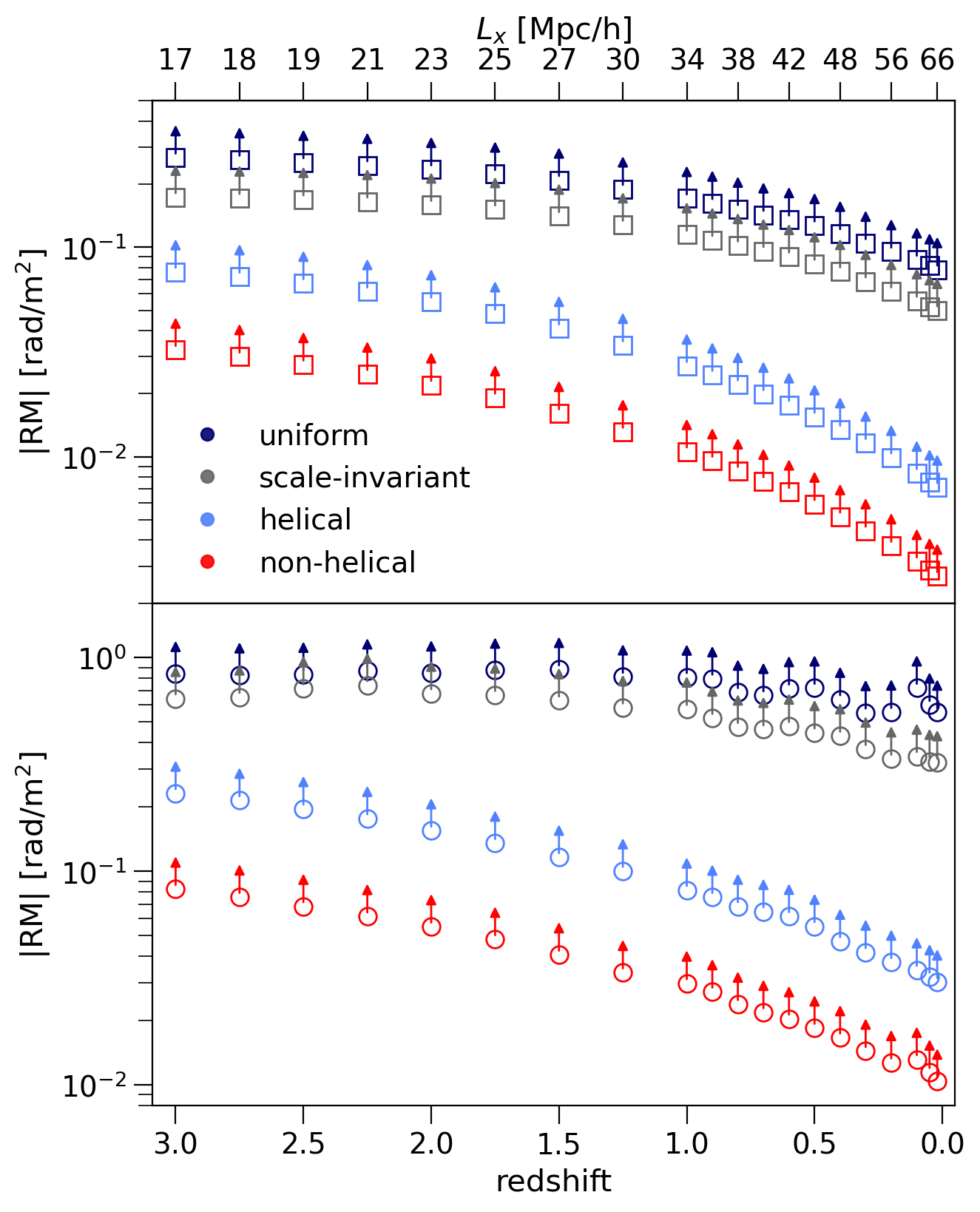}
    \includegraphics[width=5.80cm]{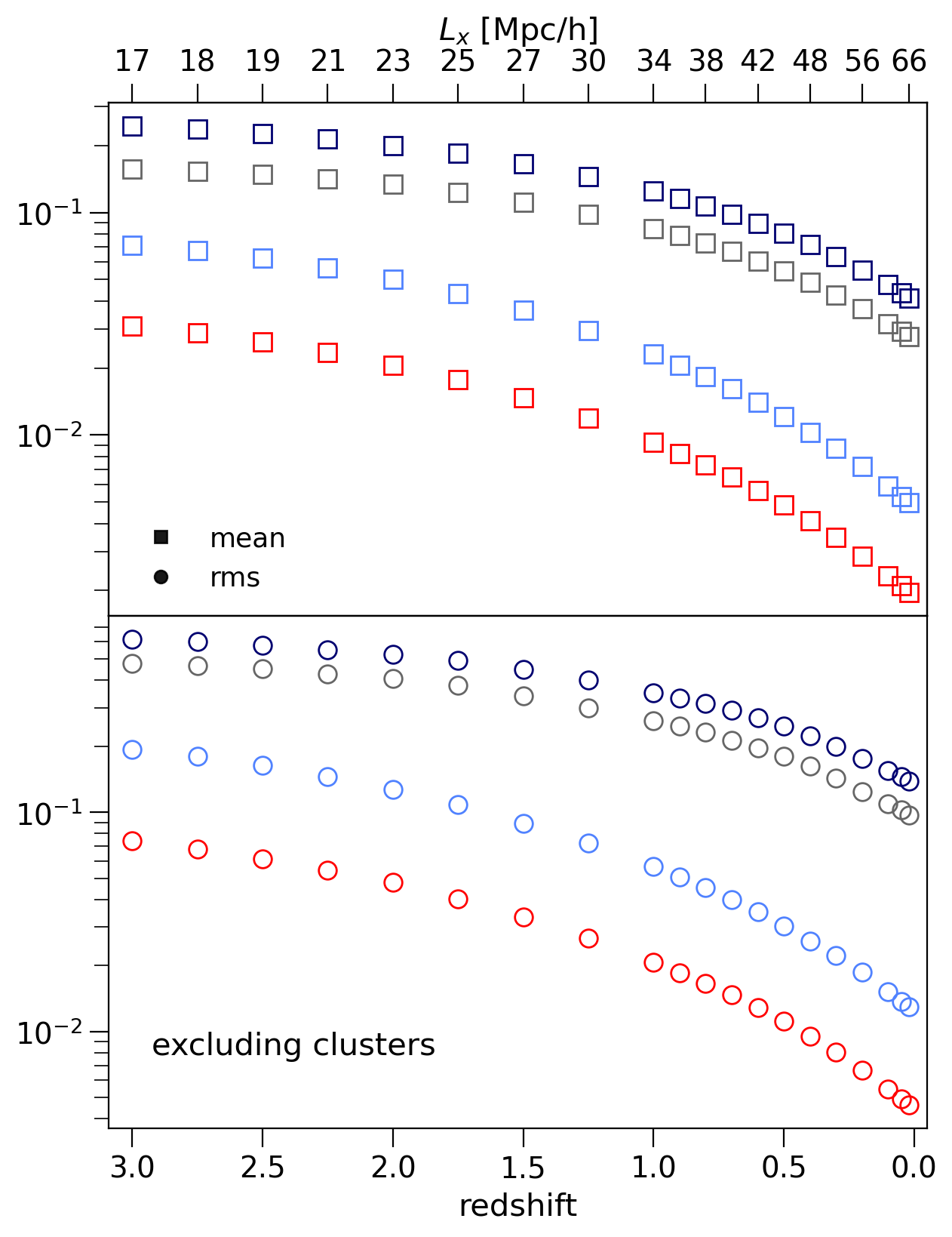}
    \includegraphics[width=5.80cm]{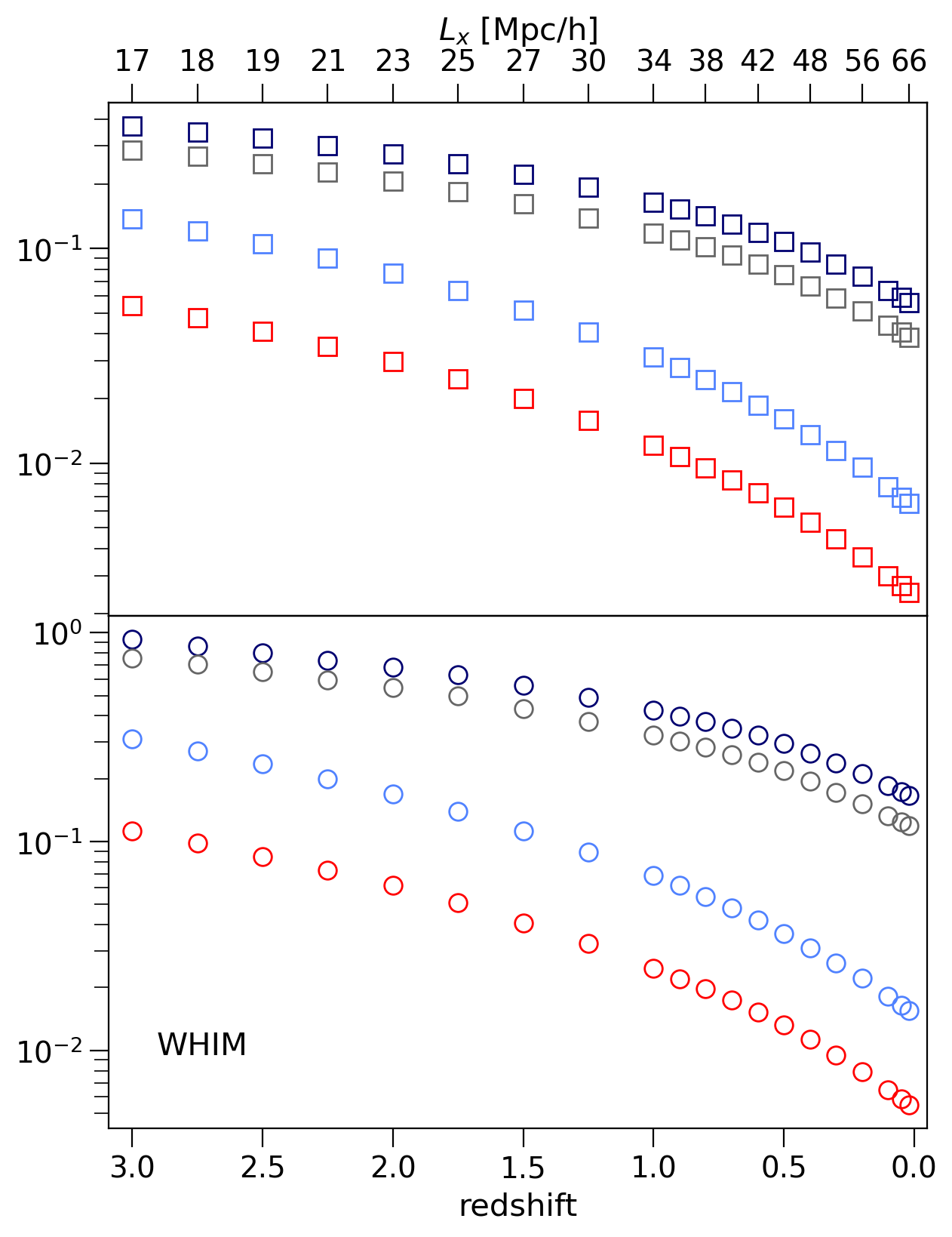}
    \caption{Redshift evolution of the mean and rms statistics of the
    absolute $|\text{RM}|$ for the whole simulating volume (first column),
    for the regions excluding galaxy clusters ($\rho /\langle \rho \rangle < 1.3 \times 10^{2}$; second column)
    and for the regions satisfying the criteria for WHIM
    ($10^5 \leq T \leq10^7$, $ \rho /\langle \rho \rangle < 1.3 \times 10^{2}$; third column).
    The upper panels show the mean values and the lower panels show the rms values.
    Statistics of all these regions exclude the lowest-density regions (satisfying the $ \rho /\langle \rho \rangle < 2 \times 10^{-2}$ criterion).}
    \label{fig:RM_regions}
\end{figure*}

In the following, we extend the analysis to a range of redshifts.
In Figure~\ref{fig:RM_regions} we show the redshift evolution of the
mean and rms statistics of $|\text{RM}|$ within
different environments and for different seeding scenarios. 
We analyzed a total of 20 cosmological boxes (corresponding to
20 redshift bins) in a redshift range of $3 \leq z \leq 0.02$.
This range is particularly relevant, for example, for the upcoming WEAVE-LOFAR survey
\citep[][]{WEAVE_LOFAR2016}, where one expects to obtain spectroscopic
redshifts for all polarized radio sources detected in the LOFAR Two-meter
Sky Survey \citep[LoTSS; ][]{Shimwelletal2019} up to $z<1$.
In the first column of Figure~\ref{fig:RM_regions}, we show the statistics from the whole simulating box 
while the second and third columns show the RM statistics in the regions excluding galaxy clusters and in the warm-hot intergalactic medium (WHIM), respectively. The regions excluding galaxy clusters imply the same criterion as in Figure~\ref{fig:RM_PDF}, while for the WHIM region we additionally set the temperature criterion $10^{5}\,$K $\lesssim T \lesssim 10^{7}\,$K.
Due to the cosmological expansion,
it is expected that the mean and rms RM will decrease with time. This is in particular true for the lowest-density regions of the cosmic web where there is almost no turbulent amplification (see the second and third columns of Figure~\ref{fig:RM_regions}). 
As we can see, the highest rms and mean $|\text{RM}|$ values are obtained
when including cluster environments and in the WHIM, which are the densest regions.
The first is consistent with previous numerical work, where it has been found
that the resulting RM is dominantly contributed by the density peaks
along the LOS \citep[e.g.,][]{AkahoriRyu2010,AkahoriRyu2011}.
Note that in these regions our reported RM values are lower than the typical observed values
due to our limited resolution (see discussion in Section \ref{sec:NumAsp}).
RMs of hundreds rad m$^{-2}$ have been observed in clusters
\citep[see][for a sample of galaxy clusters]{2016A&A...596A..22B}.
Since clusters are especially underresolved in our simulations, the RM
values in these regions should be interpreted only as lower limits.

We characterize the $|\text{RM}|_{\text{rms}}$ evolution by fitting the data to the following equation:
$|\text{RM}|_{\text{rms}}= \text{RM}_0(1+z)^{\alpha}$.
We show the fitted values for each environment for the four models in Table~\ref{tab:Tab3}. 
Since the definition of RM (see Equation~\ref{eq:RM}) includes the proper values of the electron density, magnetic field, and integration path length, we expect it to scale as RM$\sim (1+z)^2$, with $\alpha=2$. However, as we can see from Table~\ref{tab:Tab3}, $\alpha<2$ for all environments in the uniform and scale-invariant case; i.e., the decrease of RM values with redshift will be slower in these scenarios. This means that magnetic fields along with density do not only decrease in these regions, as it is expected by the expansion of the universe but are subject to amplification.

Recently, \cite{Vernstrometal2019} and \cite{OSullivanetal2020}
presented a new approach in order to isolate the extragalactic
RM variance.
The method relies on comparing pairs of extragalactic radio sources
and computing the RM difference between them. Following \cite{OSullivanetal2020}, the statistical results come from
comparing the RM from sources at the same redshift (named as physical pairs), e.g., double-lobed radio galaxies, and from sources at different redshifts (named as random pairs).
In order for us to analyze our simulations and compare to these recent
work, we would have to carry out a careful study of stacking cosmological boxes (see \citealt{ThomasCarlberg1989} and \citealt{Scaramellaetal1993}, for pioneering work) and defining light cones before integration
(see \citealt{AkahoriRyu2011} and \citealt{Vazzaetal2020}
in the context of RM).
Such a study is out of the scope of this paper and we leave it
for future work.
Nevertheless, we can give a first-order estimate on the RM difference by using the information in Figure~\ref{fig:RM_regions}. We considered the $|\mathrm{RM}|$ distribution function of the simulation box to be representative of each redshift. In this way, we select the variance of the distribution to be the representative value at each redshift.

We analyzed a total of 20 redshift bins up to $z=3$, where we obtained 190 different combinations of redshift pairs. Note that here we do not take into account the spatial distribution of possible sources within each simulation box. Instead we assume the PDF statistics to be representative of the simulation box at that redshift. 
We refer the reader to Appendix~\ref{App:RM_sources} for a word on the
distribution of sources as a function of redshift.

We compute the variance in the environment where we exclude the regions of galaxy clusters (see second column of Figure~\ref{fig:RM_regions}) and the WHIM environment (see third column of Figure~\ref{fig:RM_regions}). These two environments are less affected by our low resolution.
We are also interested in these environments because LOFAR is not expected to detect polarized sources from intervening clusters \citep[e.g.,][]{2020A&A...638A..48S}. 
As a second step, we compute the average variance between each selected redshift (random) pair, i.e. taking into account all redshift bins in between the pair. Once this is done for all the 190 pairs, we compute the rms of all the variances.
This procedure yields an rms upper limit in the WHIM of 0.7, 0.6, 0.3, and 0.2
rad m$^{-2}$ 
for the uniform, scale-invariant, nonhelical, and helical cases, respectively. The environment where we only exclude galaxy clusters gives similar values: 0.6, 0.5, 0.3, and 0.2 rad m$^{-2}$, respectively.
These RM values are marginally 
lower than the results reported at 144 MHz in \cite{OSullivanetal2020}. After analyzing the difference between physical and random pairs, the authors concluded that the excess Faraday rotation contribution between random pairs has an upper limit of 1.9 rad m$^{-2}$. This work places an upper limit of 4 nG for a PMF.
A similar analysis at 1.4 GHz and in a redshift range of $0<z<1$ was done in \cite{Vernstrometal2019}, where the authors obtained an excess contribution between random pairs of 
10.3 rad m$^{-2}$. 
In this case, the authors placed an upper limit of $\sim$37 nG.
Our rms RM values suggest that an initial magnetic seed larger than 1 nG for the four different seedings is required to meet these particular observational RM upper limits.
We can give crude estimates of the initial magnetic field strength by asking ourselves which initial magnetic field strength is needed to reach the upper limit of 1.9 rad m$^{-2}$ for all our models: the results of the uniform, spatially homogeneous field (the Mukohyama model) and the scale-invariant, inflationary field would suggest an initial magnetic field strength of $\sim 3$ nG.
This is comparable to LOFAR results and remains below the upper limits of \cite{OSullivanetal2020} and \cite{Vernstrometal2019}.
Nevertheless, this crude approach alone cannot strictly rule
out the $\sim 1$ nG normalization (see discussion below);
the phase-transitional models would allow for an even larger initial magnetic seed, namely a magnetic field strength of a primordial stochastic seed (helical or nonhelical) $\sim 6$ nG. These values
could be in agreement with CMB constraints of helical ($\sim 5.6$ nG) and nonhelical ($\sim 4.4$ nG) PMFs \citep[e.g.,][]{Plancketal2016}. Yet these high values conflict with recent Planck, Atacama Cosmology Telescope (ACT) and South Pole Telescope (SPT) constraints of a $\sim 0.05$ nG PMF \citep[see][]{galli2021}. It is important to stress that a one-to-one comparison with these recent RM observations is not strictly possible since the reported upper limits are also influenced by environmental selection effects. In this sense, we can only conclude that our first-order approximation RM analysis favors more inflationary PMFs than phase-transitional PMFs. While future simulations with a more sophisticated RM analysis could improve our predictions, future observations that will better isolate the RM signal from the diffuse WHIM and/or voids and filaments will play a decisive role in discriminating PMF models.

We expect great advances with LOFAR in the coming years. There has been significant progress in overcoming challenges in observations by using the ionospheric RM correction errors techniques \citep[e.g.,][]{2013A&A...552A..58S}. These techniques have significantly improved and will be crucial for the accurate calibration of not only the next LOFAR surveys but also for the Square Kilometre Array (SKA). Indeed, the SKA-Low and SKA-Mid \citep[][and references therein]{2015IAUGA..2252814B} is expected in coming years to provide stronger constraints on the magnetization of the universe. The increased expected number of polarized sources and knowledge of spectroscopic redshifts will enable us to make better comparisons with cosmological simulations. We expect that these advances can further help us to distinguish between the possible origins of cosmic magnetism. The present work shows that a scale-invariant and phase-transitional helical and nonhelical models still cannot be rejected. In future work, we will study a more extended range of initial conditions in parallel with the stacking technique and definition of light cones to put more stringent constrains.
\begin{deluxetable}{c c c c }
\label{tab:Tab3}
\tablecaption{
Fitted values of $|\text{RM}|_{\text{rms}}\propto \text{RM}_0(1+z)^{\alpha}$ of Figure~\ref{fig:RM_regions} for different environments of our simulations.
}
\tablehead{
\colhead{model} & \colhead{Environment} &  \colhead{$ \text{RM}_0$} & \colhead{$\alpha$} \\
 }
\startdata
{}       & Exc. clusters & $0.15$ & $1.12$ \\
Uniform  & WHIM          & $0.17$ & $1.26$  \\
{}       & All           & $0.59$ & $0.31$  \\
\hline
{}               &  Exc. clusters & $0.11$ & $1.20$ \\
Scale-invariant  &  WHIM          & $0.12$ & $1.36$ \\
{}               &  All           & $0.34$ & $0.61$ \\
\hline
{}       &  Exc. clusters & $0.01$ & $2.03$  \\
Helical  &  WHIM          & $0.02$ & $2.21$  \\
{}       &  All           & $0.03$  & $1.15$ \\
\hline
{}           & Exc. clusters & $0.01$ & $2.07$  \\
Nonhelical & WHIM          & $0.01$ & $2.22$ \\
{}           & All           & $0.01$ & $1.52$  \\
\enddata
\end{deluxetable}
%

\section{Numerical aspects}
\label{sec:NumAsp}

The spatial resolution adopted in the present simulations
place limitations on our results.
Our resolution is not sufficient to resolve, for example,
the additional magnetic amplification within galaxy clusters \citep{Xuetal2009,Vazzaetal2018,Steinwandeletal2021}. Nevertheless, in Appendix~\ref{App:ResTest} we
show that our results are robust at least on scales $\gtrsim 1\Mpc\, h^{-1}$.
We have checked the convergence of the magnetic energy power spectrum in the whole
simulation box with increasing resolution (see Figure~\ref{fig:B-PS-res}
in Appendix~\ref{App:ResTest}).
As expected, increasing the resolution results in higher power at higher wavenumbers.
On the other hand, we have also checked the trends of the magnetic field compared to the density for the
phase-transitional helical and inflationary uniform cases at different resolutions
(see Figure~\ref{fig:B-rho-res} in Appendix~\ref{App:ResTest} and corresponding discussions).
In the uniform scenario, there is no substantial change in the overall trends at higher resolution. On the other hand, the phase-transitional helical case shows subtle differences in filaments and voids, while the main differences are observed in the overdensity regions corresponding to galaxy clusters.
This seems to indicate that our results on the global properties of the filament and void regions, as well as the differences between the primordial models in those regions, are
robust for the present goal of this work.
Simulations with adaptive mesh refinement (AMR) are needed to
fully assess the discrepancies between primordial seeds within galaxy clusters. 

We also tested the dependence of our RM results on the adopted spatial resolution in the Appendix~\ref{App:ResTest}. We show the distribution of RM at $z=0.02$ at different resolutions for the uniform scenario in Figure~\ref{fig:RM_res}. 
At double resolution, the $|\text{RM}|$ values converge for the environment excluding clusters, 
while for the cluster environments we still see higher (lower) RM values at the high-end (low-end) tail of the distribution.
We expect the same follows for the stochastic scenarios. 

We used the Dedner cleaning algorithm to impose the $\nabla \cdot \mathbf{B}=0$
condition \citep[][]{Dedneretal2002}.
The main limitation of this method compared to constrained-transport (CT) schemes is the intrinsic dissipation of the scheme by cleaning
waves, which reduces the magnetic spectral bandwidth to keep
the numerical divergence under control \citep[see][]{Kritsuk2011}.
The Dedner formalism has been tested to be robust, accurate and to converge quickly to the right solution for most idealized test
problems \citep[e.g.,][]{wa09,wang10,Bryanetal2014} and for other more realistic astrophysical applications
\citep[][]{2016MNRAS.455...51H,2016MNRAS.461.1260T,2018MNRAS.476.2890B}, as long as the resolution is conveniently increased.

As mentioned in Section~\ref{sec:simulations}, we neglected
physical processes associated with radiative gas cooling, chemical
evolution, stellar and active galactic nuclei feedback.
In this way, we can solely focus on the effects of different primordial
magnetic seeding through LSS formation.
However, these unaccounted processes, are known to
pollute the rarefied regions of the cosmic web and, thus, they can potentially lower the possibility of detecting the imprints of different PMFs \citep[see, e.g.,][]{Marinnacietal2015}. We refer the reader to \citealt{Vazzaetal2017} for a comparison between the
predictions of primordial and astrophysical seeding scenarios of
magnetic fields with the \texttt{Enzo} code.

In addition, as an important caveat, we note the effect of different Riemann solvers on our results.
The most diffusive Riemann solver \citep[local Lax-Friedrichs, LLF,][]{KurganovTadmor2000} of ENZO affects
the evolution of magnetic energy in the tested nonhelical case and decreases it by factor of 2
at final redshifts (see Appendix~\ref{App:RSTest} for more details).
Consequently, the magnetic field and temperature distributions in the void regions
also show lower values at $z=0.02$ for the phase-transitional cases.
In the uniform case, on the other hand, we do not observe changes in temperature and magnetic fields due to the LLF solver.
We also checked the effect of the DEF used in our simulations, which controls thermal energy in highly supersonic bulk flows. 
We verified that the DEF does not affect the magnetic field distribution neither in the uniform case nor in the stochastic, nonhelical case. This holds for both Riemann solvers, LLF and HLL. Nevertheless, the DEF affects the temperature distribution, as expected. 
We caution the reader on the interpretation of the regions where shocks and discontinuities are created by the extreme gravitational forces. However, we do not expect these regions to be statistically significant to modify the obtained trends from our simulations.

The initial conditions used in the simulations, do not account for the
effect of magnetized perturbations on the initial matter power spectrum
(see, e.g., \citealt{Kahniashvilietal_2013_2}, \citealt{Sanatietal2020}, and \citealt{Katzetal2021} where the authors have taken this effect into account),
which would give us a self-consistent view of the cosmological initial conditions. Nevertheless, it has been recently shown that such effects will only have an impact on smaller haloes and on scales $\sim k>1\,h$ Mpc$^{-1}$ \citep[see][]{Katzetal2021}. Therefore, we would not expect significant changes in our results at the largest and most massive components of the cosmic web.

Finally, in our work we have excluded the nonideal MHD processes,
meaning that the viscous and resistive dissipation are not modeled
realistically and, therefore, the magnetic Prandtl number, i.e., the ratio
of kinematic viscosity and magnetic diffusivity, is effectively unity.
This approach is reasonable enough given the existing uncertainties and the difficulties in the characterization of galaxy clusters \citep[see, e.g.,][]{Schekochihinetal2004,Beresnyaketal2016} and larger cosmological scales. Furthermore, the ideal MHD description allows us to easily compare our work with previous work \citep[see, e.g.,][]{Alvesetal2017,Marinnacietal2018,Vazzaetal2020}. Studying higher Prandtl numbers is out of the scope of this work.

\section{Conclusions}
\label{sec:Summ}

In this work we have investigated the evolution of PMFs through the formation of LSS.
For the first time, we have compared inflationary and phase-transitional
initial seed magnetic fields with cosmological MHD simulations.
We have explored four types of initial magnetic seeds: (i) spatially
homogeneous (uniform) and (ii) statistically homogeneous (scale-invariant)
magnetic fields generated in an inflationary epoch, and (iii) helical
and (iv) nonhelical magnetic fields representing a phase-transitional
scenario.
In the latter three models the initial magnetic spectra reflect
the physics of the early universe when the magnetic seed develops
Kolmogorov-like turbulent spectra through its MHD decay, while the former case mimics a primordial magnetogenesis according to
the Mukohyama model \citep[][]{Mukohyama2016}. 

The main results of our work can be summarized as follows:

\begin{itemize}

\item \textit{The role of the initial magnetic field strength}.
A higher normalization of the initial magnetic field leads to higher
magnetic field values at later redshifts. 
However, the overall trend of the distribution of final magnetic fields in
different cosmic environments is not affected by the amplitude of the initial seed field. 
Regarding the temperature distribution, we note that phase-transitional seedings (nonhelical case) may lead to extra heating in the void regions as a result of possible turbulent decay of these fields.
In addition, the higher (initial $\geq 0.5\,$nG) magnetic field realization reveals larger differences between the inflationary and phase-transitional models. This suggest that an impact of the stronger initial seed fields will be imprinted on rarefied cosmic regions. 
Hence, both the strength and topology of the seed fields will be of notable relevance for the studies accounting for the effects of magnetic fields on the reionization history of the universe
\citep{SethiSub2005,Minodaetal2017}.

\item \textit{Traces in the cosmic web}.
Phase-transitional and inflationary scenarios lead to variations in the final magnetic field distribution of the cosmic web. The magnetic amplification in the inflationary models tends to follow the law of adiabatic gas contraction in voids (partially) and filaments, while a deviation from this law is evident in the phase-transitional models. The overall magnetization of galaxy clusters and bridges as well as of voids in the inflationary models can be orders of magnitude higher than in the phase-transitional scenarios, although the differences between the models on the galaxy clusters' and bridges' scales will be a subject of our future study (and should be confirmed with higher resolution runs). 
Discernible differences (with a lower magnitude) between the seeding scenarios are also observed in filamentary structures, where again inflationary seed fields show the largest magnetic amplification.

\item \textit{Possible inverse cascade}.
The characteristic peak of the magnetic power spectra in the phase-transitional helical and nonhelical cases shifts toward small scales at late redshifts, $z<6$. This means that, during this epoch, preferred scales due to structure formation would initially quench the energy transfer from small to large scales. At later stages, $z<3$, we observe a shift of the peak spectrum from small to large scales in the nonhelical case without an increase of correlation length, though. Therefore, our results cannot unambiguously confirm the existence of an inverse cascade.

\item \textit{Magnetic correlation length}. The final correlation length
in the inflationary seedings is larger than that in the phase-transitional
seedings (reaching $\sim 3$, $4$, $1$, and $0.8\Mpc\,h^{-1}$ in the uniform,
scale-invariant, helical, and nonhelical models, respectively).
Previous modeling of PMFs in the early universe (in the radiation-dominated epoch)
showed the same trend: inflationary, scale-invariant scenarios
lead to coherent magnetic structures on larger scales
than in the phase-transitional cases \citep{Brandenburgetal2017, Brandenburgetal2018,Mandaletal2020}.
We found that the final magnetic correlation length in the phase-transitional cases
is strongly correlated with the initial peak spectrum,
 limiting the magnetic growth at selected scales of the cosmic web.

\item \textit{Uniform versus stochastic inflationary models}. 
The late ($z>6$) evolution of spectral magnetic energy of an inflationary, scale-invariant case at scales $\lesssim 10\Mpc\,h^{-1}$ shows similar amplification as the inflationary uniform case.
A uniform, homogeneous
seed magnetic field that is customary to use in cosmological MHD simulations is a good representation of a scale-invariant magnetic field on scales smaller than $\sim 29 \Mpc\,h^{-1}$ at $z=0.02$.

\item \textit{Nonhelical versus helical phase-transitional models}.
The spectral evolution of the phase-transitional helical and nonhelical models are similar at all wavenumbers. However, the helical model exhibits more amplification as a result of larger initial correlation length 
and power on a characteristic scale than in the nonhelical scenario. We have shown that, within the limitations of our modeling, it will be hard to distinguish observationally between helical and nonhelical scenarios.

\item \textit{RM predictions}. 
Significant differences are observed
in Faraday rotation measure maps for different PMF models.
These differences arise both in the collapsed objects and in the low-density regions of the cosmic web.
We computed the rms $|\text{RM}|$ excess coming from random redshift pairs ($z\leq 3$) for the regions excluding galaxy clusters and the WHIM. RM values for all the models are lower than expected from the recent observations reported at 144 MHz \citep[][]{OSullivanetal2020}. Our RM analysis favors inflationary seed fields with larger magnetization levels in filamentary structures.

\item \textit{Non-Gaussianity}. 
We find non-Gaussian trends in the magnetic field PDFs for both inflationary and phase-transitional seedings. This is also imprinted on the distribution functions of the absolute RM, where all models show deviation from a log-normal distribution. 
The low-end tail ($10^{-12}-10^{-10}\,$G) of the magnetic field PDFs is similar for all models; on the contrary, we observe larger differences between the models in the $\text{RM}$ distribution function for both the low- ($10^{-8}-10^{-5}$ rad m$^{-2}$) and high-end tails ($10^{-1}-10^{2}$ rad m$^{-2}$) of the PDFs.
\end{itemize}

In summary, our results indicate that phase-transitional and inflationary
PMFs lead to different realizations of the magnetized 
cosmic web (retaining the information of magnetic initial conditions on the
largest scales of the universe). The differences can potentially be probed
observationally.
The Faraday rotation measures from our simulations manifest the traces
of the initial magnetic seeding. 
A stronger and more correlated RM signal is
expected from inflationary scenarios as a result of larger initial correlation lengths and higher final magnetization levels in filaments from these scenarios. 
In future work we will complement our analysis by stacking the cosmological boxes and producing light cones to give more realistic estimates of RMs.
Future observations (e.g., SKA) will detect the RM signal over a large extent of the sky and have the potential of unravelling the origin of magnetic fields on filamentary scales. 
Then the results of future work can be readily compared to those observations probing the large-/small-scale nature of the seed magnetic fields.

Finally, future numerical work related to high-energy gamma-ray propagation in cosmic voids 
(for relevant studies, see, e.g., \citealt{Dolagetal2011,Alvesetal2017,Vazzaetal2017} and \citealt{Bondarenkoetal2021} for a recent work) will be relevant to probe the seed magnetic fields studied in our work. 
It is not clear a priori what kind of topology of void magnetic fields is responsible for the suppression of the secondary gamma-ray flux \citep[][]{NeronovVovk2010}. 
Therefore, studying inflationary and phase-transitional scenarios in the context of the magnetization of cosmic voids 
could help to more stringently discriminate between the competing magnetogenesis scenarios. 
This, in turn, will help us understand the effects of 
such fields on the reionization history of the universe and first structure 
formation \citep[see, e.g.,][]{Kohetal2021}. Our work 
gives a first step and a novel approach in the search for the origin of cosmic magnetic fields.  
Future effort in combining state-of-the-art MHD cosmological simulations and more realistic initial magnetic field conditions
will be needed to
explore the role of the primordial fields on galaxy-cluster scales and down to smaller scales.

\acknowledgments
\small{
The computations described in this work were performed using the
publicly available \texttt{Enzo} code (http://enzo-project.org), which is
the product of a collaborative effort of many independent scientists from
numerous institutions around the world.  Their commitment to open science has helped make this work possible. 
We also acknowledge the \textit{yt} toolkit \citep{yt-Turk2011}, which was used as the analysis tool for our project.
This work was supported by the Norddeutscher Verbund f\"ur Hoch- und H\"ochstleistungsrechnen (HLRN)
under project number: hhp00046 with P.D.F. as principal investigator.

We thank Franco Vazza for sharing the ENZO initial setup and for his
comments on the first revision of the manuscript.
We appreciate useful discussions and comments from Emma Clarke, Klaus Dolag,
Sayan Mandal, Jens Niemeyer, Alberto Roper Pol, and Alexander Tevzadze.
S.M.\ acknowledges the financial support from Shota Rustaveli National
Science Foundation of Georgia (SRNSFG, grant No.\ PHDF\_19\_4101) and Volkswagen foundation. 
P.D.F.\ was supported by the National Research Foundation (NRF) of Korea through grant Nos. 2016R1A5A1013277 and 2020R1A2C2102800.
A.B.\ acknowledges support from the Swedish Research Council
(Vetenskapsr{\aa}det, grant 2019-04234).
M.B.\ acknowledges the financial support from the European Union's Horizon 2020 program under the ERC Starting Grant ``MAGCOW'', no.\ 714196. Finally, we thank the anonymous reviewer for his/her helpful suggestions that improved the quality of this manuscript.

{\large\em Software:} The source codes used for
the simulations of this study, ENZO \citep{2019JOSS....4.1636B} and
the {\sc Pencil Code} \citep{JOSS}, are freely available
on  \url{https://github.com/enzo-project/enzo-dev}
and \url{https://github.com/pencil-code/}. 

\section{Data Availability}
The data underlying this article will be shared on reasonable request to the corresponding author.
}


\bibliography{PMFs}{}
\bibliographystyle{aasjournal}

\appendix
\section{{\sc Pencil Code} Initial Conditions}
\label{AppA}
We use different initial magnetic fields in our four sets of simulations.
First, we use a uniform magnetic field, which was already implemented in the \texttt{Enzo} code.
Second, a stochastic magnetic field requires a special treatment as it is
the result of an MHD simulations with the {\sc Pencil Code} \citep{JOSS}.
In this code, we solve the compressible resistive MHD equations \citep{Brandenburgetal_1996},
in which we advanced the magnetic vector potential.
We adopt an ultrarelativistic equation of state, appropriate during
the radiation era.
The initial condition consists of a Gaussian-distributed spatially
random field such that the magnetic energy spectrum has a certain shape:
proportional to $k^{-1}$ for the scale-invariant spectrum and proportional
to $k^4$ for wavenumbers $k<k_*$ and proportional to $k^{-5/3}$ for $k>k_*$
in the case of a phase-transitional initial magnetic field.
The simulation domain is triply periodic and normalized such that its
volume is $(2\pi)^3$, and so the smallest wavenumber in the domain is
then $k_1=1$.
We choose $k_*=120$ for the runs with phase-transitional initial magnetic fields.
The sound speed, $c_{\rm s}$ and the mean density are normalized to unity.
The rms magnetic field, and thus the initial rms Alfv\'en speed is
$v_{\rm A}=0.1...0.2$.
We solve the equations for a time interval $\Delta t$ such that
$c_{\rm s} k_1 \Delta t=10$, corresponding to
$v_{\rm A} k_* \Delta t\approx200$.

We generate 
the magnetic field components as eigenfunctions of the curl operator, but 
for helical fields the sign of the eigenvalue is the same for all wavevectors;
see Equation~(23) of \cite{Brandenburgetal2017}.
Helical magnetic fields are still random and isotropic, but they have
the same systematic swirl of field lines everywhere in the domain, just
like a box with randomly oriented screws which all have the same sense
of winding.

The use of a stochastic initial magnetic field requires a modification
of \texttt{Enzo} where the field is normalized such that for each
component $B_{s,i}$, the input (proper) magnetic field is:
\begin{equation}
    B_{s,i} = \frac{B_{P}}{\sigma_{P}}B_{{\rm com},i}\,(1 + z_{\rm ini})^2,
\end{equation}
where $B_{P}$ is the input from the {\sc Pencil Code}, $\sigma_{P}$ is
its standard deviation, and $z_{\rm ini}$ is the initial redshift ($z=50$).
We set $B_{{\rm com},i}= 1 \text{nG}$ and take $B_{u,i} = \text{RMS}[B_{s,i}]$
for the $B$-field components in the case of a uniform initial field.
In this way, we normalize all the initial conditions to have the same
magnetic energy, $\int (B^2/8\pi) \, dV$.

\section{Resolution tests}

\label{App:ResTest}
\begin{figure}[htbp]
    \centering
    \includegraphics[width=8.2cm]{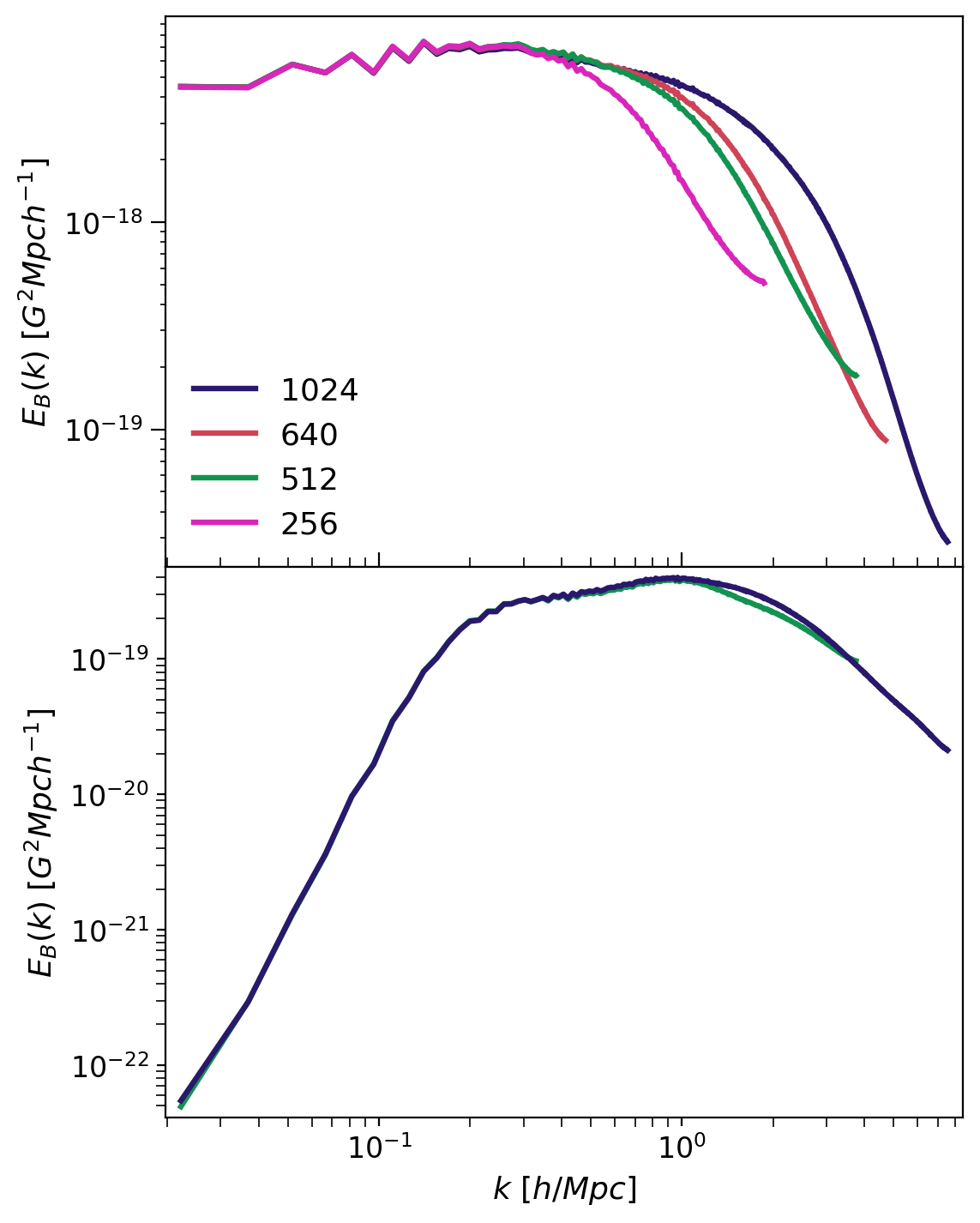}
    \caption{Magnetic energy power spectrum with increasing resolutions
    (comoving: 264 $h^{-1}$kpc, 132 $h^{-1}$kpc, 105 $h^{-1}$kpc, 66 $h^{-1}$kpc)
    for the uniform (top panel) and helical cases
    (bottom panel) at $z=0.02$.}
    \label{fig:B-PS-res}
\end{figure}
\begin{figure}[htbp]
    \centering
    \includegraphics[width=8.2cm]{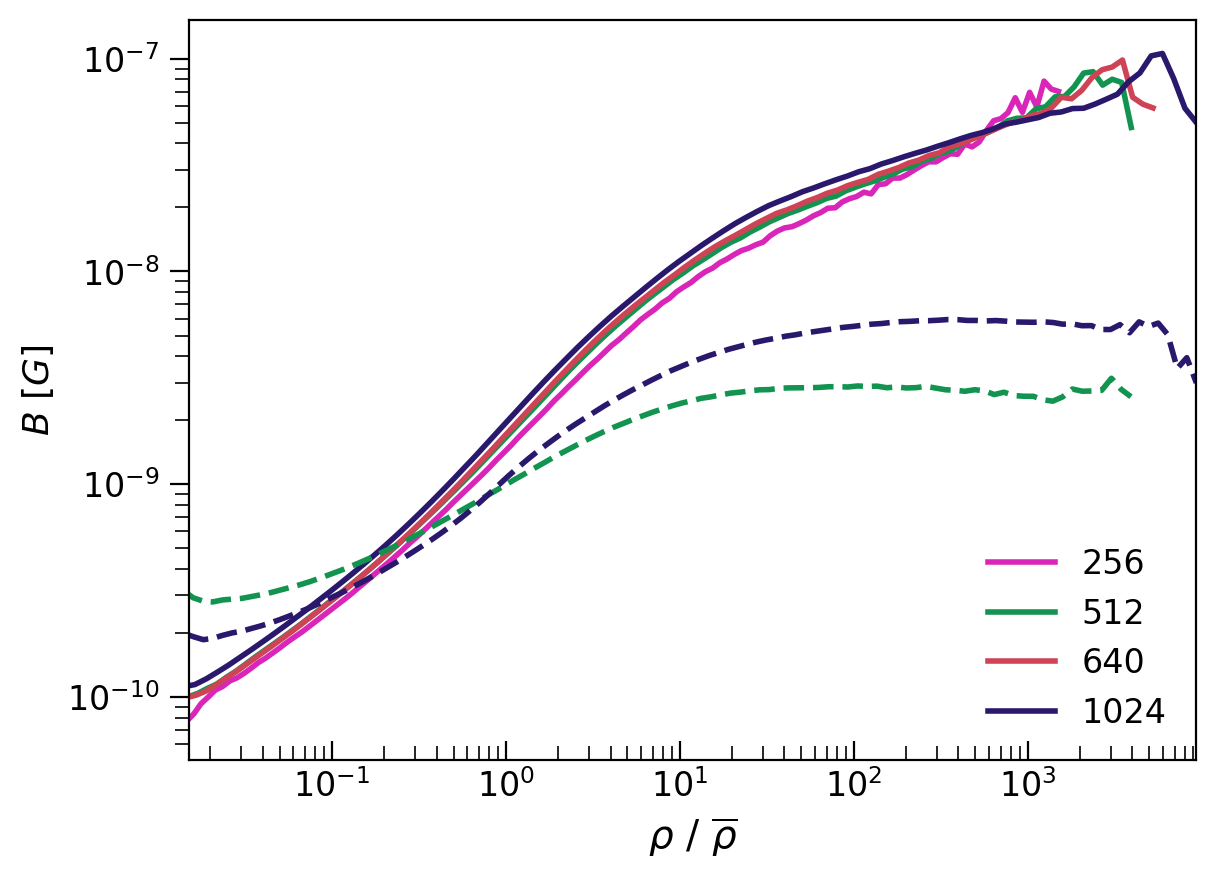}
    \caption{
    The median magnetic field versus overdensity profile for increasing grid points ($256, 512, 640, 1024$) at $z=0.02$. Solid lines: uniform seeding case; dashed lines helical case.
    } 
    \label{fig:B-rho-res}
\end{figure}
\begin{figure}[htbp]
    \centering
    \includegraphics[width=8.3cm]{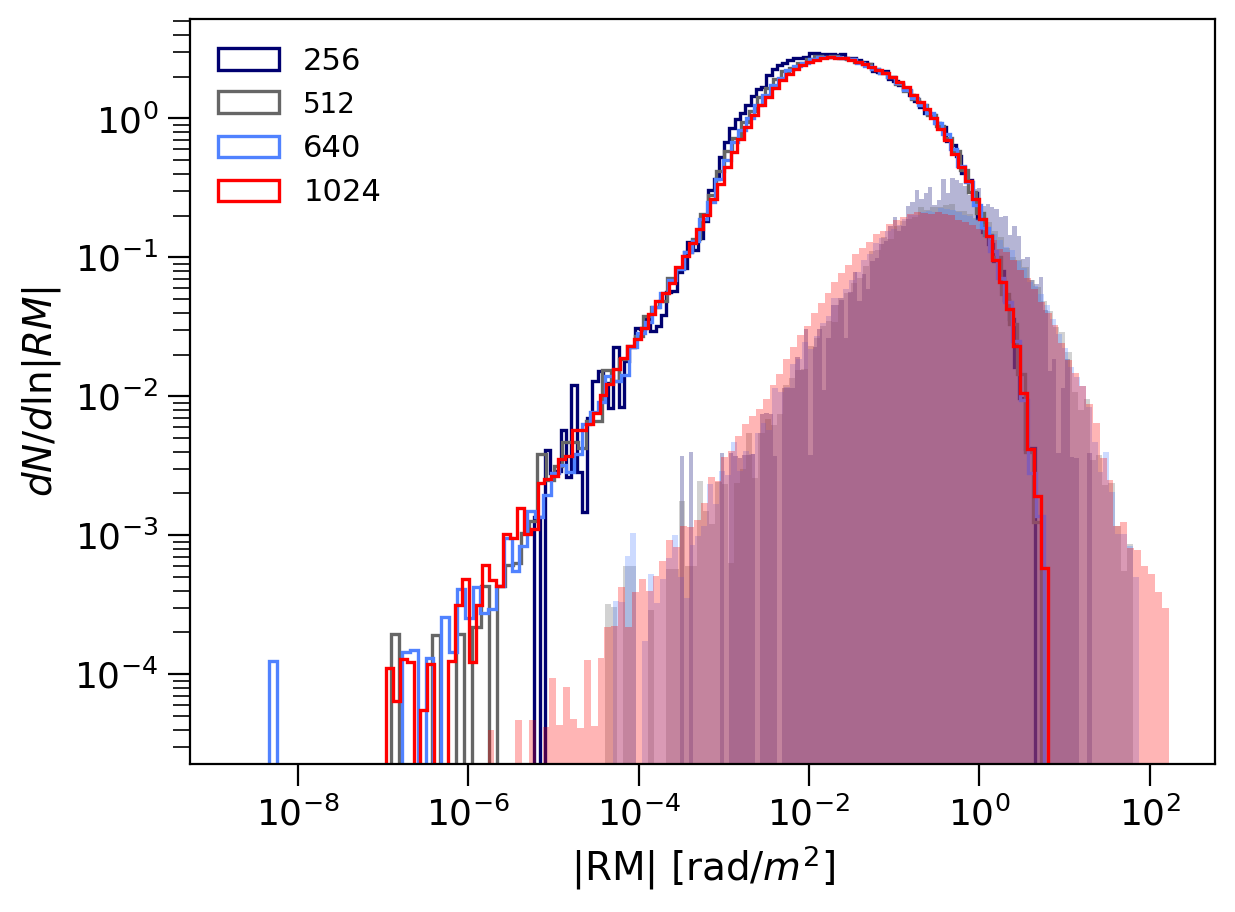}
    \caption{$|\text{RM}|$ distribution function dependence on resolution (comoving: 264 $h^{-1}$kpc, 132 $h^{-1}$kpc, 105 $h^{-1}$kpc, 66 $h^{-1}$kpc) at $z=0.02$ for the uniform seeding. The solid lines represent regions excluding galaxy clusters, while the shaded areas show distributions only for galaxy clusters.
    } 
    \label{fig:RM_res}
\end{figure}
In order to test the convergence of our simulations, we compared our
setup with the higher-resolution runs including  uniform and helical magnetic 
seeding only (as the evolution of the power spectrum seems to be similar toward smaller scales 
in the uniform and scale-invariant cases and helical and nonhelical cases). 
These runs used the same set of cosmological parameters and the box
size as is used in the main paper. Since the initial stochastic magnetic field distributions are obtained
from the {\sc Pencil Code}, in the helical case we used 
the AMR technique of the \texttt{Enzo} code
\citep[][]{Bryanetal2014} to reach the same resolution ($132h^{-1}\kpc$) as in the uniform case with 
$1024^3$ grid points. In Figure~\ref{fig:B-PS-res} we show
the growth of the magnetic field power spectrum as a function of spatial
resolution.
As we can see, for our case (i.e., the run with $512^3$ grid points,
$\Delta x=132 h^{-1}\kpc$ resolution), the magnetic energy in Fourier space for the uniform seeding (solid lines)
would be underestimated on scales  $\sim k>0.5\,h\Mpc^{-1}$ (cluster scales), while in the helical case (dashed lines) 
only the minor difference is expected when doubling the resolution. As we can also see the larger scales
 $k\la0.5\,h\Mpc^{-1}$ are well converged in
our simulations.

Conversely, we can see in Figure~\ref{fig:B-rho-res}
that the median profiles show no differences in the trends when the
higher-resolution simulation is seeded by the uniform magnetic field.
On the other hand, in the case of helical
seeding (dashed lines), higher magnetic field
strength is achieved in the filaments' and clusters' regions and lower strengths in the voids for the higher-resolution run. Therefore, we would
expect the differences between the inflationary and phase-transitional
models to be decreased in these regions as a result of higher
resolution.

In Figure~\ref{fig:RM_res} we also show the resolution dependence of the PDF of absolute RM. 
We see that RM values show convergence in the regions where we exclude galaxy clusters, while RMs in galaxy clusters are mostly affected at the high-end and low-end tail of the distribution.
This means that our analysis of Sec.~\ref{sec:RM_analysis} will be more affected in galaxy-cluster regions. 
\begin{figure}[t!]
    \centering
    \includegraphics[width=8.2cm]{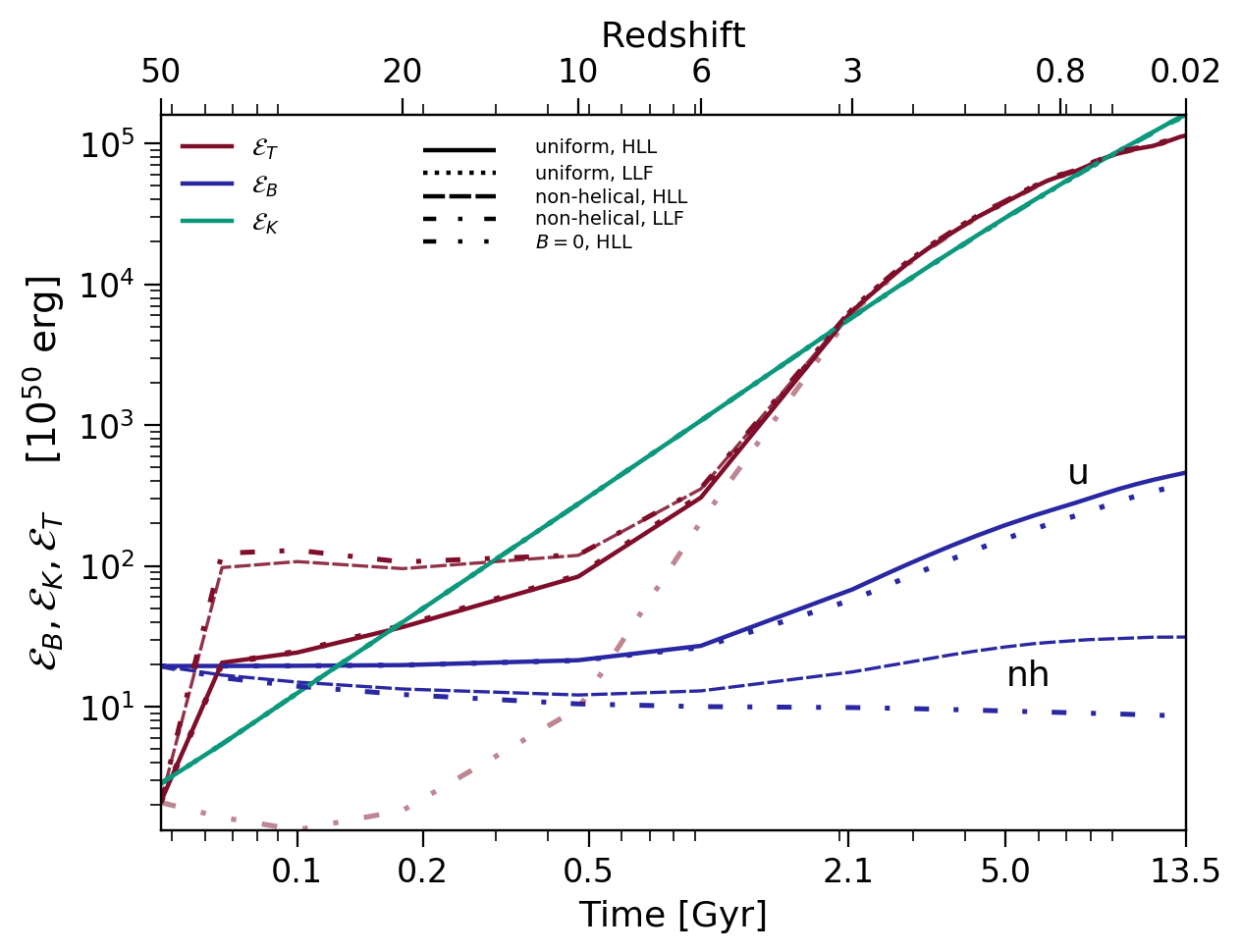}
  \caption{Evolution of magnetic, thermal, and kinetic energies for the uniform and nonhelical cases when using HLL (solid, dashed, dashed-dotted lines, where the latter corresponds to the $B=0$ case) and LLF (dotted and dashed-dotted lines) Riemann solver schemes.
    } 
    \label{fig:EnEv_LLF}
\end{figure}
%

\section{Testing the Riemann solver}
\label{App:RSTest}

An important aspect of our work is also to determine the dependencies of our results on the adopted numerical methods. 
As discussed in Section \ref{sec:general}, PMFs are expected to heat the IGM at high redshifts. Nevertheless, the absence of heating and cooling physics in our simulations poses a challenge for the interpretation of large thermal energies observed in the turbulent 
helical and nonhelical models 
at redshifts $z>10$ (see Figure~\ref{fig:EnergyEv}).
In order to check the energy evolution, we have tested our setup with
two different Riemann solvers, the LLF and the HLL.
Note that the LLF scheme is considerably more diffusive than the HLL scheme.
In Figure \ref{fig:EnEv_LLF} we show the evolution of thermal, kinetic, and magnetic energies obtained from these two setups.
Additionally, we show the evolution of thermal and kinetic energies
for a $B=0$ case using the HLL Riemann solver.
As we can see, the pure hydrodynamical setup shows the lowest thermal
energy at high redshifts in comparison to the MHD cases.
On the other hand, the evolutionary trends of the thermal and kinetic
energies in the MHD cases are not affected by the change of Riemann solvers.
This shows that the already observed differences between turbulent,
nonhelical and uniform models in thermal energy (see discussion in
Section \ref{sec:general}) are not dependent on the selection of the
Riemann solver.
However, the LLF scheme substantially affects the evolution of the
magnetic energy in the nonhelical case.
The stronger numerical diffusion completely suppresses the magnetic field amplification 
at lower redshifts, suggesting that this method may not be optimal for the study of stochastic PMFs.

\begin{figure}[htbp]
\begin{center}
\includegraphics[width=0.45\textwidth]{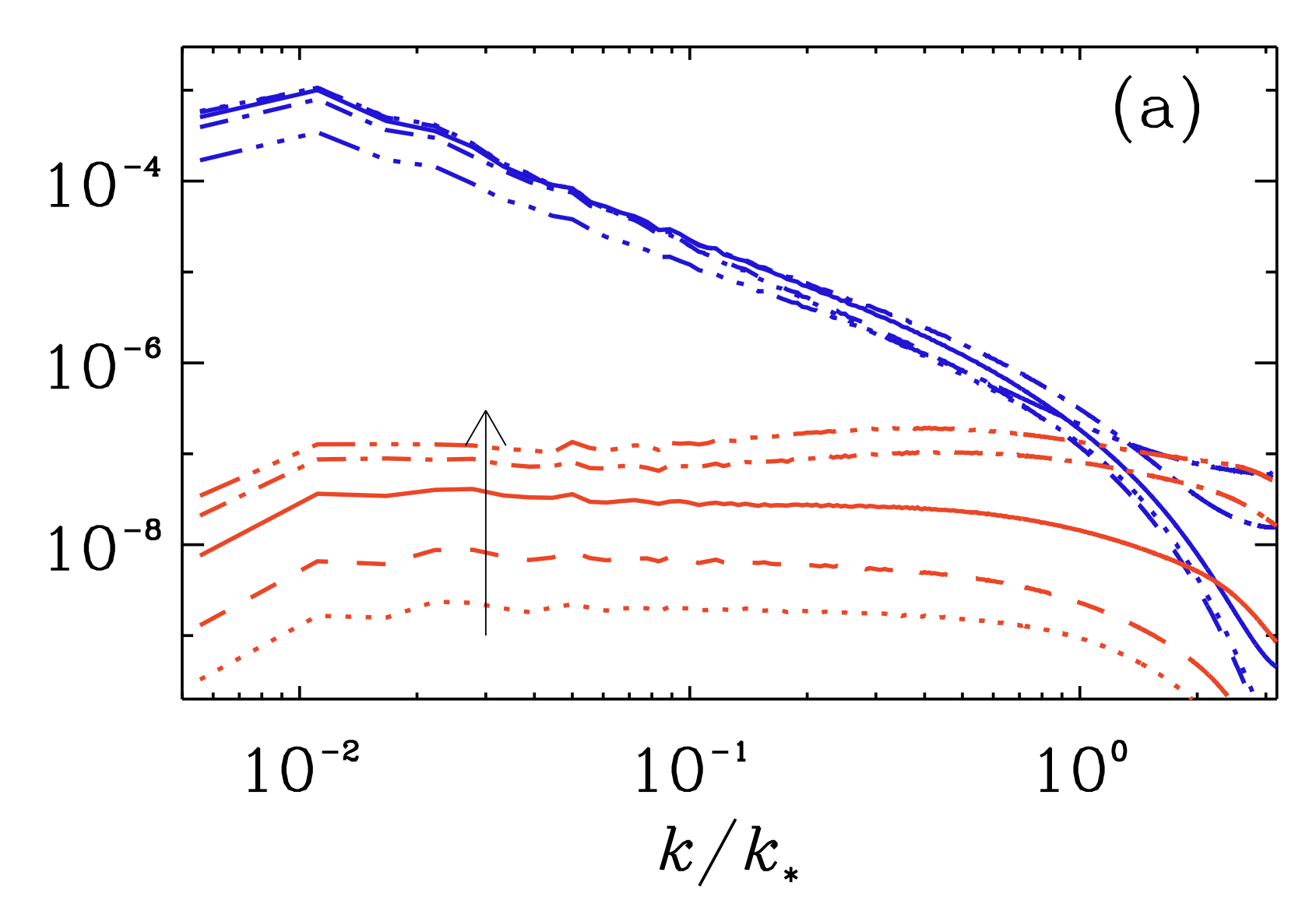}
\includegraphics[width=0.46\textwidth]{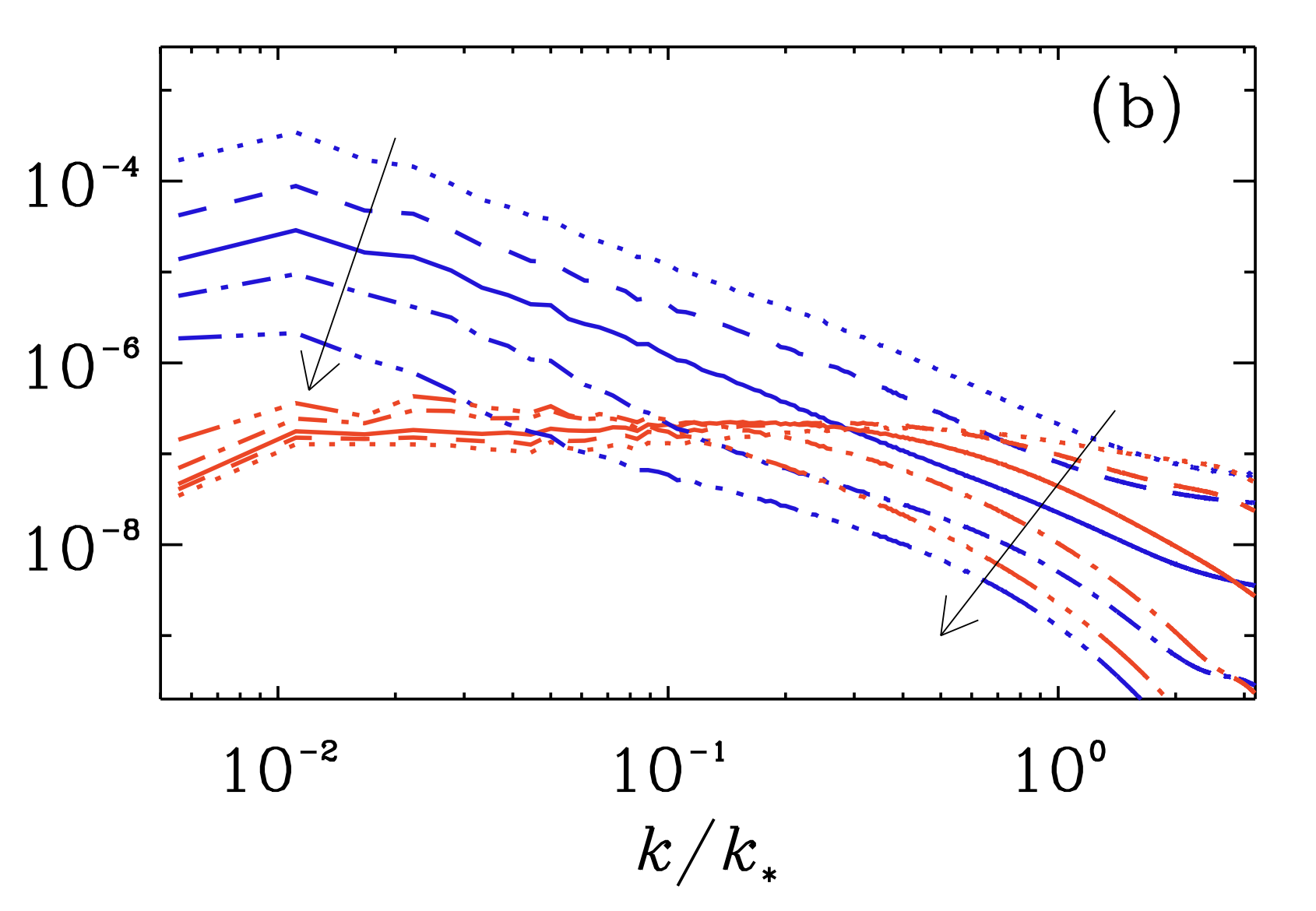}
\end{center}
\caption{
MHD simulations with an initial kinetic energy spectrum proportional
to $k^{-2}$ in the presence of a weak homogeneous magnetic field.
Panels (a) and (b) show the early and late evolution for magnetic
energy spectra (red) and kinetic energy spectra (blue).
}
\label{pkt1152_2panels_I1152EKm2d}
\end{figure}

\begin{figure}[ht!]
    \centering
    \includegraphics[width=8.5cm]{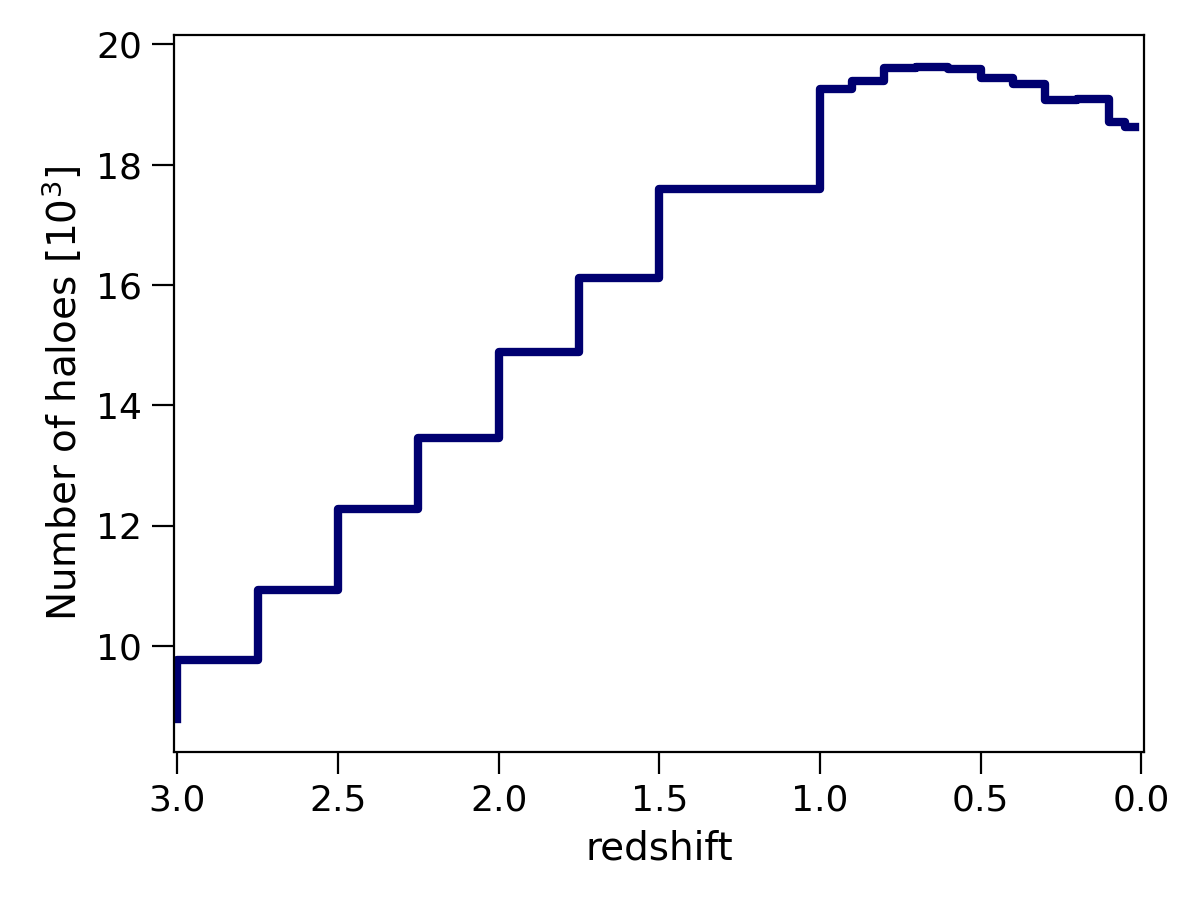}
    \caption{Number of virialized objects with baryonic mass range
    $10^7 M_{\odot} \leq M \leq 10^{10} M_{\odot}$
    as function of redshift.
    The masses have been calculated within the sphere enclosing the
    virial radius of each source.
    This radius is an output of the \textit{yt} halo finder.
    }
    \label{fig:num_sources}
\end{figure}

\section{Tangling of a homogeneous field}
\label{App:Tangling}

In \Sec{subsec:MagPS}, we noted the rather different response to an
imposed (uniform) magnetic field in the present simulations (see the
first panel of Figure~\ref{fig:B-PS}) and those of \cite{Mandaletal2020},
where magnetic fields at the scale of the domain began to grow only
after smaller-scale magnetic fields have grown first.
An important difference, however, is that our present cosmological
simulations always possess a large-scale velocity field, which was
not present in the simulations of \cite{Mandaletal2020}.
The large-scale velocity field in the cosmological simulations leads to
tangling of the uniform imposed magnetic field, which leads to the
instantaneous growth immediately at the beginning of the simulation.

To verify our reasoning above, we now repeat the simulations presented in
Figure~5 of \cite{Mandaletal2020} with a weak imposed field, but now with
a random initial velocity field with an initial kinetic energy spectrum
proportional to $k^{-2}$.
The result is shown in Figure~\ref{pkt1152_2panels_I1152EKm2d}, where
we show the resulting magnetic and kinetic energy spectra.
Note the increase of spectral magnetic energy at all wavenumbers already occurs from early times onwards.
This is caused by the tangling of the homogeneous magnetic field by the
initial large-scale velocity field with a $k^{-2}$ spectrum.
This therefore confirms our reasoning in \Sec{subsec:MagPS} concerning
the rather different response to an imposed magnetic field in the present
simulations and those of \cite{Mandaletal2020}.

\section{RM sources}
\label{App:RM_sources}
As we mentioned in Sec.~\ref{sec:RM_analysis} our statistical RM analysis does not take into account the spatial distribution of sources at each redshift.
Nevertheless, we show the expected redshift distribution of sources in Figure~\ref{fig:num_sources} for completeness. 
For this Figure, we used the \textit{yt} halo finder \citep{SkoryTurk2011}, which identifies the groups of linked DM particles based on the \cite{EisHu1998ApJ} algorithm, for each redshift in order to find a selected number of haloes with a baryonic mass range, $10^7 M_{\odot} \leq M \leq 10^{10} M_{\odot}$.
This range resembles the masses of FRII galaxies, which are the typical type of polarized sources found in large catalogs \citep[e.g.,][]{2009ApJ...702.1230T,2018A&A...613A..58V}.

\listofchanges
\end{document}